%% file: main.tex
\documentclass[letterpaper,twocolumn,10pt,hyphens]{article}
\usepackage{usenix2019_v3}

\usepackage{graphicx}
\usepackage{times}
\usepackage{epsfig}
\usepackage[tight]{subfigure}
\usepackage{setspace}
\usepackage{amssymb}
\usepackage{array}
\usepackage{comment}
\usepackage[english]{babel}
\usepackage{amsmath}
\usepackage{tabularx}
\usepackage{balance}
\usepackage{tikz}
\usepackage{cleveref}
\usepackage{breqn}
\usepackage{multirow}
\usepackage{listings}
\usepackage{booktabs}
\usepackage{colortbl}
\usepackage{pifont}

\crefformat{section}{\S#2#1#3}

\newcommand{\cmark}{\ding{51}}
\newcommand{\xmark}{\ding{55}}

\newcommand{\code}[1]{{\fontfamily{txtt}\selectfont {#1}}}

\newcommand{\MyCaption}[1]{
  \vspace{-1.3ex}
  \caption{#1}
  \vspace{-1.6ex}
}

\newcommand{\MyTableCaption}[1]{
  \vspace{-0.2ex}
  \caption{#1}
  \vspace{-1ex}
}

\newcommand{\MySection}[1]{
  \vspace{-1ex}
  \section{#1}
  \vspace{-1ex}
}

\newcommand{\MySubsection}[1]{
  \vspace{-1.2ex}
  \subsection{#1}
  \vspace{-0.8ex}
}

\begin{document}

\date{}

\title{\Large \bf Arcadia: A Fast and Reliable Persistent Memory Replicated Log}

\author{
\rm Shashank Gugnani\thanks{Now at Oracle}\\
The Ohio State University
\and
{\rm Scott Guthridge}\\
IBM Research
\and
{\rm Frank Schmuck}\\
IBM Research
\and
{\rm Owen Anderson}\\
IBM Research
\and
{\rm Deepavali Bhagwat\thanks{Now at Google}}\\
IBM Research
\and
{\rm Xiaoyi Lu}\\
UC Merced
}

\maketitle

\begin{abstract}
The latency and bandwidth properties of byte-addressable persistent memory (PMEM) have the potential to significantly improve system performance over a wide spectrum of applications.  But at the same time, PMEM brings considerable new challenges for the programmer compared with earlier technologies.  In particular, PMEM provides only 8-byte write atomicity, can flush writes out of order, and its availability is always limited by node failures.  It's possible to work with the atomicity and ordering constraints of PMEM directly by carefully sequencing the order of store operations and inserting explicit flush and fence operations at each ordering point.  But this is tedious and error-prone: too many flush operations defeat the performance benefits of PMEM, and even with generous use, it is difficult to prove that a given program is crash-consistent. Logging is a great abstraction to deal with these issues but prior work on PMEM logging has not successfully hidden the idiosyncrasies of PMEM.  Moreover, shortcomings in the log interface and design have prevented attainment of full PMEM performance.  We believe that a log design that hides the idiosyncrasies from programmers while delivering full performance is key to success.  In this paper, we present the design and implementation of Arcadia, a generic replicated log on PMEM to address these problems.  Arcadia handles atomicity, integrity, and replication of log records to reduce programmer burden.  Our design has several novel aspects including concurrent log writes with in-order commit, atomicity and integrity primitives for local and remote PMEM writes, and a frequency-based log force policy for providing low overhead persistence with guaranteed bounded loss of uncommitted records.  Our evaluation shows that Arcadia outperforms state-of-the-art PMEM logs, such as PMDK's libpmemlog, FLEX, and Query Fresh by several times while providing stronger log record durability guarantees.  We expect Arcadia to become the leading off-the-shelf PMEM log design.

\end{abstract}

\input{Texts/intro}
\input{Texts/motivation}
\input{Texts/primitives}
\input{Texts/design}
\input{Texts/evaluation}
\input{Texts/related}
\input{Texts/conclusion}





\bibliographystyle{plain}
\bibliography{Bibs/bibfile}

\end{document}

%% file: Texts/intro.tex
\MySection{Introduction}
Persistent Memory (PMEM) provides byte-addressable access to data and persistence across power cycles with latency and bandwidth properties that make it an attractive storage technology.  At nearly the speed of DRAM, it promises significant performance gains for a number of important workloads. Unfortunately, the persistence of PMEM is hard to realize in practice~\cite{pmem-blog,nalli2017analysis}.  Until store operations have been flushed from CPU caches, they remain volatile, thus are lost on power failure.  Moreover, the CPU can reorder stores and implicitly evict cache lines at any time, making it necessary for the programmer to meticulously order stores and explicitly flush cache lines in order to maintain crash consistency.

Even then, the hardware provides only 8-byte atomicity for local writes and so far \emph{no} atomicity for remote writes over the network, further complicating the work of updating data in a crash-consistent way~\cite{pmem-atomicity}.  Using the memory interface for speed also makes PMEM intrinsically \emph{local}, thus node failures and network partitions curtail its availability.  This is in stark contrast with the 4KB sector atomicity typical of other common block-based storage technologies and the ability to share storage across a storage area network such as NVMe-oF.

A further complication not unique to PMEM is that uncorrectable media errors or DIMM failures may result in permanent loss of data~\cite{zhang2019pangolin}.

All of these problems can be solved by introducing a replicated log.  First, logging solves the problem of atomicity of updates for crash consistency across power failures.  Before making changes to PMEM, the programmer first records into the log a description of the changes to be made (typically redo or undo records), then makes the changes to primary storage in PMEM. If the system crashes while making the changes to PMEM, log recovery at the next start-up restores consistency of the data in PMEM by completing (with redo) or reverting (with undo) any partial updates~\cite{Mohan1992ARIESAT}. Second, replicating the log across nodes solves the problems of node failures, network partitions, uncorrectable media errors, and PMEM DIMM failures.

To preserve the performance benefits of PMEM as fast primary storage, the log itself must also reside in fast PMEM storage.  The log must be implemented with performance and reliability in mind and must be able to handle all the programming challenges of PMEM.  To meet these, a lot of work has to go into the design of the log.  But once this has been done, the log provides a platform on which the programmer can easily realize the benefits of PMEM as primary storage with full performance and reliability.

In addition to serving as a log for PMEM as primary strorage, the fast PMEM log can also provide crash consistency and excellent update latency for primary storage on any other persistent medium, reducing the log latency that often limits update rate.

PMEM's byte addressability is an excellent match with remote direct memory access (RDMA).  With RDMA operations, data can be transferred quickly to and from PMEM on a remote node.  Using RDMA with PMEM is not without complications, however, because it suffers from many of the same atomicity and persistence problems as local PMEM.  Thus, the abstractions the log uses to manage the challenges of PMEM locally must be extended to PMEM over RDMA.

\noindent \textbf{Why doesn't prior work solve all problems?}
Prior work on PMEM logging~\cite{taleb2018tailwind,pmdk,xu2019finding,wang2017query,huang2014nvram,kim2016nvwal,arulraj2016write,yang2020filemr} places inordinate trust in hardware to retain and atomically update data. Prior works were designed with the assumption that PMEM has reliability similar to DRAM. However, reliability for PMEM is a more complex and underappreciated problem\footnote{With DRAM, if there is data corruption or a media error, simply restarting the application solves the problem, but this is not possible with PMEM.}: media errors and software bugs can \emph{permanently} corrupt data, and the reliability of each memory-cell degrades as it is used, potentially leading to premature failure~\cite{zhang2019pangolin,pmem-rel}. So, in practice, without sufficient redundancy and integrity checking, data may be lost or corrupted, putting additional burden on the programmer to achieve reliability.  Further, prior work does not adequately provide write concurrency necessary for full PMEM performance while also ensuring in-order commit (monotonicity).

In this paper, we present the design and implementation of \textbf{Arcadia}\footnote{Named after the Greek province of Arcadia.}, a generic replicated log to tame the challenges of PMEM.  We make the following novel contributions in this paper:

\begin{enumerate}
\item A generic interface for Arcadia that has two distinguishing features: it hides the complexities of replication and persistence, and it decouples steps that must be serialized from those allowing concurrency. Log append concurrency provides for application processing overlap, hides the latencies of checksum calculation and replication, and provides concurrency necessary for full performance.
\item Persistence, atomicity, and integrity primitives for both local and remote accesses to PMEM. These abstractions hide the weak consistency of PMEM, ensure reliable persistence, and make PMEM easier to use.
\item A design and implementation using the primitives above to handle atomicity, integrity, and replication of log records with total-order semantics.
\item A frequency-based log force policy for providing low overhead persistence with guarantees on bounded data loss. 
\end{enumerate}
We deploy Arcadia on two servers with Intel Optane PMEM that are connected with RDMA. Our analysis shows that Arcadia significantly outperforms state-of-the-art PMEM logs, such as FLEX~\cite{xu2019finding}, PMDK's libpmemlog~\cite{pmdk}, and Query Fresh~\cite{wang2017query} while providing stronger log record durability guarantees. We expect Arcadia to be used as an off-the-shelf log implementation for any storage system using logging, in particular systems that have weak consistency or disaggregated storage.


%% file: Texts/motivation.tex
\MySection{Background and Motivation}

In this section, we discuss the background and motivation of our work.

\MySubsection{Persistent Memory}

PMEM technologies, such as PCM~\cite{lee2010phase,lee2009architecting}, ReRAM~\cite{Akinaga2010ResistiveRA}, STT-RAM~\cite{tulapurkar2005spin}, 3D-XPoint~\cite{hady2017platform}, and memristor~\cite{memristor} are disruptive technologies in storage system design.  With three orders of magnitude better latency and an order of magnitude better bandwidth compared to flash~\cite{mittal2016survey,yang-fast20}, PMEM is an excellent candidate for storing latency critical data.  Given that PMEM is an emerging technology with much higher cost than flash~\cite{kim2014evaluating}, we do not expect PMEM to replace flash as primary storage in the foreseeable future.  Rather, we expect it to be used in the role of latency critical logging and metadata storage in next generation systems.

Despite its performance benefits, the write atomicity and ordering problems of PMEM make it difficult to achieve both performance and crash consistency.



\MySubsection{RDMA Networking}

Modern networking interconnects such as InfiniBand~\cite{ib} and RDMA over Converged Ethernet (RoCE)~\cite{roce2} are heavily used in high-performance computing to achieve high throughput and low latency. These technologies are now being increasingly adopted in data centers worldwide.
Through RDMA, a process can read or update memory of a remote process while minimizing remote involvement. Data transmission bypasses both the local and remote OS, allowing zero-copy operation with data transferred directly between network and memory controllers.
Protocol processing can generally be offloaded to the hardware, further improving performance.
RDMA provides both two-sided (Send and Recv) and one-sided (Read and Write) primitives. The one-sided primitives benefit from being able to transfer data without any critical path involvement of the remote side.  Choosing the appropriate primitive is necessary for high-performance and low latency communication.

\MySubsection{PMEM Persistence over RDMA}

As with DRAM, RDMA can be used to directly access data in PMEM. Unfortunately, the one-sided RDMA primitives were not designed for the complicated persistence properties of PMEM. RDMA writes to PMEM are acknowledged as soon as data has reached the remote NIC. So, there is no guarantee that data has been written to the desired memory location as it could be cached within the NIC or PCIe buffers. Once data has been transferred out of the NIC, there is still no guarantee that it has been made persistent because modern NICs can put data directly into volatile CPU caches, still outside of the PMEM persistence domain. As a result of the complicated write path, there is no persistence or atomicity guarantees for RDMA writes to PMEM. This makes the design of a replicated log challenging because a fundamental requirement of logging is to make record updates atomic and persistent. One solution for solving both atomicity and persistence issues is to leverage two-sided RDMA primitives in a request and response replication model with the remote side responsible for persisting data before responding to the replication request~\cite{snia1}. In doing so, the performance benefits of one-sided RDMA cannot be realized. Another solution is to leverage special hardware primitives, such as RDMA Commit~\cite{rcommit} or to always force the NIC to write data to PMEM instead of CPU caches. The former relies on hardware support from the NIC that is currently not available in any commodity product, while the latter requires BIOS configuration changes that also penalize the performance of any network traffic not destined for PMEM. Therefore, one-sided RDMA writes cannot directly be leveraged to replicate data to remote PMEM unless both the persistence and atomicity limitations are addressed.

\MySubsection{Limitations of Prior Work}


\begin{table}[t]
\centering
\setlength\arrayrulewidth{0.72pt}
\resizebox{0.48\textwidth}{!}{
\begin{tabular}{l|cccc}
\toprule
\textbf{Log} & \textbf{Device/Node} & \textbf{Network} & \textbf{Media} & \textbf{Power} \\
\textbf{Design} & \textbf{Failure} & \textbf{Partition} & \textbf{Error} & \textbf{Loss} \\ \midrule
\rowcolor{lightgray}
PMDK~\cite{pmdk} & \xmark & \xmark & \xmark & \cmark \\
FLEX~\cite{xu2019finding} & \xmark & \xmark & \xmark & \cmark \\
\rowcolor{lightgray}
Query Fresh~\cite{wang2017query} & \cmark & \cmark & \xmark & \cmark \\
Tailwind~\cite{taleb2018tailwind} & \cmark & \cmark & \cmark & \xmark \\
\rowcolor{lightgray}
\textbf{Arcadia} & \cmark & \cmark & \cmark & \cmark \\
\bottomrule
\end{tabular}
}
\MyTableCaption{Comparison of resilience to key failure scenarios with related work. Note that Arcadia is the only system that is robust to all of these failure scenarios. In addition, it is the only one to provide concurrent writes with in-order commit.}
\label{tab:related}
\end{table}

The performance critical nature of logging has led many researchers to propose new log implementations on PMEM as a way of improving end-to-end system performance. Table~\ref{tab:related} presents a comparison of resilience to key failure scenarios between Arcadia and related work. Several works have looked at designing both unreplicated and replicated logging protocols on PMEM~\cite{taleb2018tailwind,pmdk,xu2019finding,wang2017query,huang2014nvram,kim2016nvwal,arulraj2016write,yang2020filemr,zhang2015mojim}. Unreplicated protocols, like PMDK's libpmemlog~\cite{pmdk} and FLEX~\cite{xu2019finding} are by design not resilient to node or device failures. This makes it hard to apply them in systems with reliable and replicated primary storage. The log is a single point of failure in persistent systems, so its resilience must be at least that of primary storage; therefore, replicated logging is essential for a robust storage system. Unfortunately, prior work on replicated logging places more trust in PMEM hardware's ability to retain data than is prudent.  This is because these works were designed with the assumption that PMEM has similar
reliability to DRAM. However, reliability for PMEM is more complex: media errors and software bugs can permanently corrupt data, and the reliability of each memory-cell degrades as it is used, potentially leading to premature failure~\cite{zhang2019pangolin,pmem-rel}. So, in practice, without sufficient redundancy and integrity checking, data may be lost or corrupted, pushing the burden of reliability onto the programmer. For instance, Query Fresh~\cite{wang2017query} and Mojim~\cite{zhang2015mojim} do not have any mechanism to handle memory corruption (resulting from software accidentally writing data to the wrong place) or undetected media errors, so it is possible for clients to read silently corrupted data. Further, Mojim does not offer concurrent log writes. Tailwind~\cite{taleb2018tailwind} assumes the presence of DMA capable battery-backed DRAM buffers to guarantee persistence. Such special hardware support is impractical in a production system and we are not aware of any commodity system offering such support.

Providing concurrency for log writes on PMEM is another dimension where prior work falls short. It is challenging to provide concurrent writes to the log because multi-threaded concurrency is non-deterministic and storage systems typically require log writes to preserve a total order (monotonicity)  to guarantee correctness. Without a total order, consistency of systems cannot be guaranteed on recovery. While some prior work does provide concurrent log writes, those designs are for block-based disk or flash storage, not for byte-addressable PMEM. PMEM makes this problem more challenging because cache lines may be evicted implicitly at any time. So, if log data is accessed concurrently and in-place, there are no ordering guarantees for data persistence. Prior work completely sidesteps this issue by fully isolating log writers to preserve the total log order but at the cost of limited or no concurrency.

Arcadia was designed to overcome these shortcomings of prior work. Through replication, Arcadia can survive device/node failures and network partitions. Further, by leveraging novel integrity and atomicity primitives for PMEM, Arcadia is resilient to memory corruption, media errors, and power loss. Arcadia also provides concurrency while preserving in-order commits. To achieve this goal, Arcadia separates log writes into a set of distinct steps and \emph{only} isolates those that require serialization. In this manner, unnecessary synchronization can be avoided and concurrency is maximized.

%% file: Texts/primitives.tex
\MySection{Primitives for Local and Remote Accesses}

Handling atomicity, integrity, and persistence of accesses to PMEM is a non-trivial task. In this section, we first present primitives for durable local and remote writes to PMEM. Building upon these, we present primitives for atomically and reliably accessing data in local or remote PMEM. Crucially, these primitives handle the complexities of PMEM + RDMA persistence and hide the weak and complicated persistence properties of PMEM. We distinguish the primitives as Persistence, Replication, Integrity and Atomicity.

\noindent \textbf{Use Cases.} The proposed primitives help make PMEM easier to use. The integrity primitive can be used to reliably write data once, or when the data being overwritten is known not to be needed anymore. The atomicity primitive can be used when updating existing data in place, where we need to make sure that if there is a failure during write, the old data remains readable. We show how these primitives can be applied in Arcadia to achieve the persistence, integrity, and atomicity requirements of log writes. The persistence and replication primitives are building blocks for the integrity and atomicity primitives, which Arcadia uses to format different portions of the log -- the integrity primitive for log records and the atomicity primitive for the log header. Additional details on how Arcadia uses these primitives can be found in~\cref{sec:works}.

\noindent \textbf{Persistence Primitive.} To guarantee persistence locally, we construct a primitive using hardware access macros. It takes a location and length in PMEM and makes sure that data previously stored at that location is persistent. It relies on architecture-specific instructions such as \code{clwb} and \code{sfence} that can guarantee that data has been flushed from the volatile CPU caches into PMEM. 

\noindent \textbf{Replication Primitive.} Our goal is to construct a replication primitive which can work on currently available commodity hardware and does not require hardware-modification or BIOS configuration changes. To this end, we design a single round-trip protocol for both replicating and persisting data to remote PMEM. We use the hybrid RDMA-Write-with-Immediate (RDMA-Write-Imm) operation to both replicate data and signal the remote server to force data to storage using the persistence primitive. The `immediate' value in this operation carries with it the length of the data to force while its starting address is obtained from the RDMA-Write completion event. Essentially, the RDMA-Write-Imm acts as an asynchronous RPC to signal the remote server to persist data. Once the persist operation is complete, the remote server sends an acknowledgement using a (two-sided) RDMA Send. The local server can use this acknowledgement as an assurance of remote data persistence. This approach preserves the asynchrony of one-sided RDMA while also guaranteeing remote persistence.

\begin{figure}[t]
\centering
\includegraphics[width=0.357\textwidth]{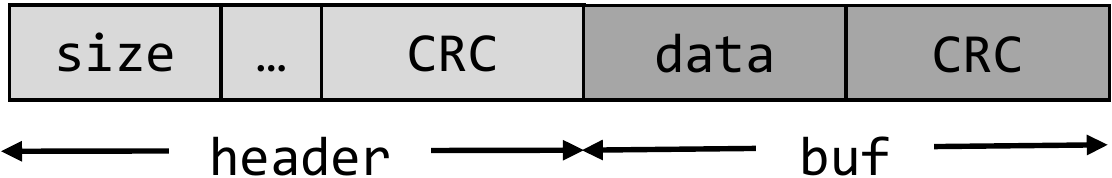}
\MyCaption{Integrity Primitive Data Layout in PMEM}
\label{fig:integrity}
\end{figure}

\begin{lstlisting}[label={lst:integrity},
language=Python,
basicstyle=\fontfamily{txtt}\selectfont\footnotesize,
numbers=left,
stepnumber=1,
frame=lines,
morekeywords={function,this,true,false},
keywordstyle=\fontfamily{txtt}\selectfont\footnotesize\color{blue}\bfseries,
commentstyle=\footnotesize\color{gray},
float=t,
caption={Integrity Primitive}]
# header and buf are preallocated in PMEM
function ReliableWrite(data, size)
    header = generateHeader(data, size)
    header.crc = crc32(header) # Header CRC
    memcpy(buf.data, data, size) # Copy data
    buf.crc = crc32(data) # Data CRC
    # Replicate + Force header and data
    return rdma_write_and_force(header, buf)

function ReliableRead(data)
    rdma_read(header, buf) # Remote read
    if (header.crc != crc32(header))
        return false
    if (buf.crc != crc32(buf.data))
        return false
    memcpy(data, buf.data, header.size)
    return true
\end{lstlisting}

\noindent \textbf{Integrity Primitive.} This primitive is designed for reliably reading and writing data in PMEM (e.g., log records). Reliability here means that we should be able to verify the integrity of data. This requires protection against two situations -- the first where writes may be torn, i.e., only part of data is persisted, and the second where data is corrupted because of undetected media errors. To achieve reliability, we add a header field to every data item with checksums protecting both of header and data. Figure~\ref{fig:integrity} provides an overview of the data layout for this primitive. The header contains the size of the data buffer and can be used to store additional information, for instance, to identify the type of data stored. Listing~\ref{lst:integrity} presents a pseudo-code of the primitive. The \code{rdma\_write\_and\_force()} function represents the replication primitive and can be replaced by the persistence primitive if replication is not desired. For reliably writing data, checksums (such as CRC32) for header and data are generated. By protecting both using checksums, there are no ordering requirements of writes or persistence barriers for the header and data. In addition, there is no requirement on atomicity of writes to PMEM. This is crucial because replicating data using one-sided RDMA does not provide any atomicity guarantees. In our approach, a single replicate and persist operation is sufficient for the entire write. When reading data, both the header and data checksum must be checked before data can be safely copied. Header integrity must be validated before data integrity, otherwise the size field in the header may not be correct. As an optimization in some circumstances, the header checksum can be replaced by a special integrity check value, such as a log sequence number (LSN), to eliminate the cost of generating the checksum.

\begin{figure}[t]
\centering
\includegraphics[width=0.34\textwidth]{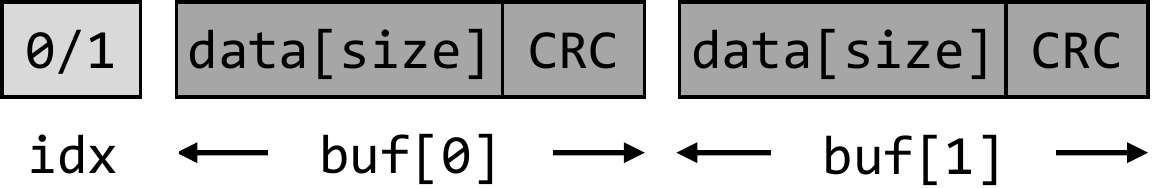}
\MyCaption{Atomicity Primitive Data Layout in PMEM}
\label{fig:atomicity}
\end{figure}

\begin{lstlisting}[label={lst:atomicity},
language=Python,
basicstyle=\fontfamily{txtt}\selectfont\footnotesize,
numbers=left,
stepnumber=1,
frame=lines,
morekeywords={function,this,true,false},
keywordstyle=\fontfamily{txtt}\selectfont\footnotesize\color{blue}\bfseries,
commentstyle=\footnotesize\color{gray},
float=t,
caption={Atomicity Primitive}][t]
# idx and buf[] are preallocated in PMEM
function AtomicWrite(data)
    memcpy(buf[!idx].data, data, size)
    buf[!idx].crc = crc32(data) # Generate CRC
    # Replicate + Force data and CRC
    rdma_write_and_force(buf[!idx])
    idx = !idx # Flip index
    # Replicate + Force index
    return rdma_write_and_force(idx)

function AtomicRead(data)
    # Remote read
    rdma_read(idx, buf)
    if (buf[idx].crc != crc32(buf[idx].data))
        return false
    memcpy(data, buf[idx].data, size)
    return true
\end{lstlisting}

\noindent \textbf{Atomicity Primitive.} This primitive is designed for atomically accessing data in PMEM and is required when updating an object with a fixed location (e.g., the log's header). It guarantees that the entire data is updated atomically and that data integrity can be validated on reads. Atomically updating data is challenging because writes may be torn, so data may be only updated partially. Data cannot be updated in-place because there is no guarantee of atomicity for remote writes to PMEM. To solve this issue, we propose the use of copy-on-write (CoW) for updates to data. Figure~\ref{fig:atomicity} provides an overview of the data layout and Listing~\ref{lst:atomicity} presents a pseudo-code of the primitive. We use two buffers to implement the CoW approach and switch between the two for every update. The index flag identifies the current valid buffer for reads. Checksums protect data integrity. For any update, data must be written to the invalid buffer. Once data and its checksum have been written and persisted, the index can be updated and persisted accordingly. As an optimization, to eliminate the cost of persisting the index flag, the flag can be placed in volatile storage as long as the valid data buffer can be identified on recovery using the contents of the data. Another possible optimization, to reduce space usage, is to always write to a new dynamically allocated buffer and discard the old buffer once the write has successfully completed.

%% file: Texts/design.tex
\MySection{Design}


In this section, we present the design and implementation of Arcadia. We design Arcadia as a replicated log with a single-primary, multi-backup model. It has a single multi-threaded writer doing updates (logger) and a single reader during recovery.

\MySubsection{Design Overview}

\begin{figure}[t]
\centering
\includegraphics[width=0.47\textwidth]{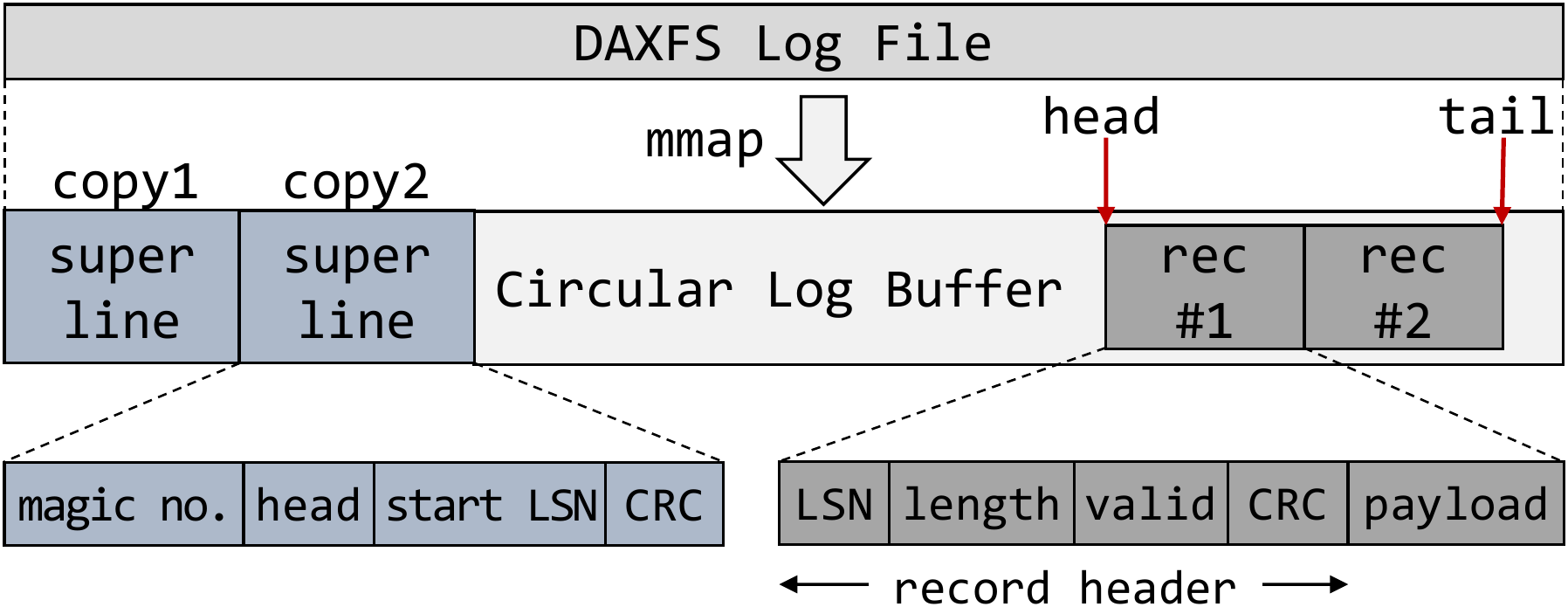}
\vspace{-0.5ex}
\MyCaption{Arcadia's Layout on PMEM}
\label{fig:layout}
\end{figure}

\noindent \textbf{PMEM Layout.} We place the entire log as a single file on a DAX filesystem formatted on PMEM. The entire file is \code{mmap}ed into the address space of the process to enable load/store access to log records. Figure~\ref{fig:layout} shows an overview of the log layout on PMEM. The log itself consists of a circular buffer of fixed size to hold the log records and a header containing information necessary to both identify and recover instances of the log. We call this header the \emph{superline}, a term combining superblock with cache line. The superline stores the head and start LSN of the log, which recovery uses to find the first valid record. We do not include the log tail in the superline to avoid the overhead of having to update it on each log append. Instead, we find the log tail on recovery by iterating over all valid records to locate the end of the log. Each log record in Arcadia consists of a header and payload pair. The header contains the length and checksum of the payload as well as a monotonically increasing LSN and a valid flag.
Arcadia relies on special values and basic consistency checks like magic numbers, LSNs, checksums, and valid flags to validate the log's content in persistent storage.

\noindent \textbf{Operation Modes.} Arcadia can be deployed in one of three operation modes -- \emph{local}, \emph{local+remote}, or \emph{remote only}. In the \emph{local} mode, log data is placed and accessed only on PMEM available on the local server. In \emph{local+remote} mode, the local copy is the primary replica of the log with one or more secondary servers serving as backups. Finally, in \emph{remote only} mode, all durable copies of the log reside on one or more remote backups, while the client only holds a volatile copy of the log.

Whenever Arcadia is configured in a mode with replication, at most one replica is local and the rest must be remote.  Because remote replicas have higher latency than local due to network overhead, it follows that when replication is in use, write latency is always limited by the network, thus having a local replica does not significantly improve performance.  A positive consequence of this is that log clients can be located anywhere there is good RDMA connectivity, and are not required to have local PMEM.


\begin{table}[t]
\centering
\setlength\arrayrulewidth{0.72pt}
\resizebox{0.45\textwidth}{!}{
\begin{tabular}{ll}
\toprule
\textbf{API Function} & \textbf{Description} \\ \midrule
\rowcolor{lightgray}
\emph{[id, ptr] reserve(size)} & Reserve space for a record \\
\rowcolor{lightgray}
\emph{int copy(id, data, size)} & Copy data into a record \\
\rowcolor{lightgray}
\emph{int complete(id)} & Mark a record as completed \\
\rowcolor{lightgray}
\emph{int force(id, freq=1)} & Force a record to PMEM \\ \midrule
\emph{id append(data, size)} & Append a record \\ \midrule
\rowcolor{lightgray}
\emph{LSN getLSN(id)} & Get record LSN \\  \midrule
\emph{int cleanup(id)} & Reclaim record space \\
\emph{int cleanupAll()} & Reclaim entire log \\ \midrule
\rowcolor{lightgray}
\emph{ITERATOR begin/end()} & Recovery iterator \\
\bottomrule
\end{tabular}
}
\MyTableCaption{Arcadia's Interface}
\label{tab:api}
\end{table}

\noindent \textbf{Interface.}
Arcadia's interface was designed with three goals -- (1) minimizing unintended synchronization, (2) providing direct access to PMEM, and (3) enabling use in any situation requiring logging. An overview of the interface is presented in Table~\ref{tab:api}.

Conventionally, log interfaces only include an \emph{append} method for log writes. Bundling the entire write as a single append method limits overlap of application compute with log writes and data persistence. To achieve the first goal, we propose a set of fine-grained methods to write data in the log. These methods break the append method into four separate stages -- reserve, copy, complete, and force. This separation provides users flexibility to place the calls at the best possible location to maximize concurrency of application and log-related activity. A single append method is also provided as a combination of the four fine-grained methods wrapped up into a single convenient operation when the fine-grained interface isn't needed.

The second goal is achieved by providing a direct pointer to the PMEM region allocated for a record using reserve. The benefit of having a direct pointer is that it allows the user to assemble the log record contents directly in PMEM without need to first build it in DRAM and then copy it. It avoids the need for a data copy in cases where the data to be logged is not already sitting in DRAM.

Finally, to achieve the third goal, we provide several generic methods to add records, reclaim space used by records, and iterate over valid records for recovery. These methods are described in detail in~\cref{sec:works}.

\MySubsection{Replication and Recovery Protocol}
\label{sec:rep_and_rec}

Arcadia implements a quorum-based protocol for replication and recovery. The write quorum ($W$) is a configurable parameter that can be adjusted based on desired performance and fault-tolerance. The read quorum ($R$) is automatically selected based on W and the number of durable replicas ($N$) to satisfy quorum requirements ($R + W > N$). Therefore, the system can tolerate up to $N - W$ replica failures when writing log records.

Arcadia assumes that it is running within an existing cluster infrastructure (such as Apache Zookeeper~\cite{apache-zookeeper}) that manages membership and quorum of nodes, and that assigns an active \emph{primary} node to control each instance of the Arcadia log and run the user application.

\noindent \textbf{Replication.}
When log data needs to be replicated, RDMA Writes are initiated to all backups (remote log replicas) in parallel. Next, we wait for the persistence acknowledgement from all backups. Once sufficient writes have completed to meet the write quorum, the data can be considered to be durably replicated. Network partitions and backup failures can disrupt this process. To handle such cases, Arcadia uses a timeout-based approach. If the backup write times out, we consider the backup to be failed and immediately close the connection with that server. This also avoids the problem of an inconsistent backup in cases with a transient network partition. As long as the number of failed backups does not impact the write quorum (i.e., $W$ backups are in complete sync), the replication process is successful. In situations where write quorum cannot be achieved, replication will fail, and as a consequence any call to \emph{force} will also fail. We delegate responsibility for these failures to the application. An easy fix that the application can use in such cases is to gracefully shutdown, add new backup servers (by copying the PMEM log files), and restart the application.

\noindent \textbf{Recovery.} 
Once the existing cluster infrastructure has assigned an active node to manage the Arcadia log, Arcadia checks that a sufficient number of copies of the log are available and consistent to meet the log read quorum, $R$. If the read quorum is not met because backups are unavailable, Arcadia fails recovery and lets the user retry after more backups come online. If  read quorum fails because too many copies are corrupted, then Arcadia cannot guarantee data consistency and recovery cannot proceed safely.
During the recovery process, we use data in the superline to detect which copies of the log are consistent and contain the most recent data. Before accepting new requests, Arcadia repairs any inconsistent or failed log copies using the consistent copies. Only inconsistent copies are modified during recovery, therefore, the recovery process is idempotent and invulnerable to repeated failures. Once recovery has completed, the primary and all backups are up to date, and new requests can be safely accepted.

\noindent \textbf{Handling Primary Failure.}
The primary node hosting the user application can fail at any time. We ensure that Arcadia handles such failures by using the cluster infrastructure. On primary failure, this infrastructure triggers leader election, selects a new primary, and informs all backups of the primary change. All backups immediately close their connection with the primary to ensure that it is fenced off. This prevents the old primary from continuing to replicate log records after it comes back up. The user application is then migrated to the new primary and restarted once log recovery is complete. Since Arcadia replicates log records synchronously, the new primary is guaranteed to have the latest log records that were forced by the client. This preserves both application consistency and correctness.

\noindent \textbf{Handling Diverging Histories.}
Diverging histories can happen due to repeated failures of different backups. For example, assume we have 3 log replicas ($A$, $B$, and $C$) with $R$=2 and $W$=2. Replica A writes a record of value $X$ at LSN 1, and then crashes. Recovery reads replicas $B$ and $C$, which have not written anything at LSN 1. They are consistent and recovery proceeds without $A$. Next, replicas $B$ and $C$ write a record of value $Y$ at LSN 1, and then crash. Recovery reads replicas $A$ and $B$ and discovers one replica with $X$ at LSN $1$, and one with $Y$ at LSN 1, creating a conflict during recovery. To solve this problem, we add an epoch field to the superline in each log copy. The epoch values start out at $1$. On each recovery, all available copies are read and the maximum epoch value is calculated. At least read quorum of copies must be readable, or recovery cannot continue. Recovery increments the epoch by 1 and writes the new value to all available copies. Writes to a write quorum of nodes must be successful to continue. Only log copies with the maximum epoch are considered valid. So, in the example above, only replicas $B$ and $C$ will be considered valid during the last recovery, thereby avoiding diverging histories.

\MySubsection{Using the Log}
\label{sec:works}

In this sub-section, we describe how Arcadia accesses and updates the log in detail. Arcadia uses the proposed integrity primitive for accessing log records and the atomicity primitive for accessing the superline. As an optimization to the integrity primitive, we use the LSN for validating the header rather than a checksum. For optimizing the atomicity primitive, we keep the index in volatile memory and identify the valid superline on recovery based on which copy has the latest start LSN.

\noindent \textbf{Concurrent Log Writes and Monotonicity.} 
Arcadia allows concurrent writes despite having in-order commits. We describe how each write method is handled internally to enable this.
The \textbf{reserve} method allocates space for log records and returns a direct pointer to the allocated space in PMEM\footnote{When configured without local PMEM, reserve returns a pointer to a buffer in DRAM.}. It is also responsible for generating the LSN for the record. To ensure monotonicity of LSNs, this method synchronously allocates log space and generates LSNs.
The \textbf{copy} method is a convenience method to copy data into a reserved record.
It uses non-temporal stores that bypass the CPU caches to copy data to PMEM efficiently. This method can be safely called multiple times to copy different data chunks into the record. Note that users may choose not to use this method and instead use \code{memcpy} or CPU stores to copy data themselves. The \textbf{complete} method is used once all data has been written to the record. Complete generates and writes the checksum for the record payload and also sets the valid flag in the record header. The \textbf{force} method guarantees that a record is durable, i.e., it will be available after a crash. It does this by replicating and persisting records. Importantly, it ensures that there are no gaps in the stream of committed records. To do this, force waits, if necessary, until all previous records have been marked valid (complete) and forces them first. Doing so ensures that records are always persisted in-order despite concurrent log writers.

The key insight here is that reserve and force are the only synchronous methods required during a log write. Multiple threads can safely call copy and complete in parallel. Although data is concurrently written into the log, the reserve and force methods carefully track the status of all records to ensure that monotonicity is still maintained.

\noindent \textbf{Log Space Reclamation.} Space reclamation is important when records are no longer necessary for recovery because the changes they describe have already been made durable. Depending on the type of logging algorithm used, records may be invalidated at different times. However, a cleanup method operating at the granularity of a record is sufficiently generic to be applicable in all cases. This method is responsible for unsetting the valid flag of the record. It also checks if there is a contiguous section of the log starting from the head (the oldest valid record) that can be reclaimed. If so, it advances the log head and updates its value in the superline. A cleanupAll method is also provided to reclaim the entire log. This method simply re-initializes the entire log and updates the superline accordingly, but preserves the current log epoch number.

\noindent \textbf{Recovery Iterator.} This iterator is used to access all valid records upon recovery from a crash. It is useful in bringing the system back to a consistent state using data in the log. Before it can be used, the recovery protocol (see~\cref{sec:rep_and_rec}) is triggered to repair any corruption and ensure consistency of all copies. The recovery iterator operates on the local log copy and obtains the log head from its superline to find the oldest valid record. To validate records, three integrity checks are performed -- (1) all LSNs leading up to the record and of the record itself must be monotonically increasing, (2) the valid flag must be set, and (3) the checksum of the payload must match. The iterator ends when any one of these checks fails. These integrity checks guarantee that any partially written, uncommitted, or corrupted records are not applied during recovery.

\MySubsection{Force Policy}

Up until now, we have assumed that log writers would always do an explicit force on the log after writing a log record, i.e. that the thread doing the logging would not be able to continue until the log record was safely committed. But doing a force operation on the log after every write is expensive.
The high cost of persisting data and the coarse granularity of writes for traditional block storage motivated researchers to propose techniques to reduce the number of synchronous, small writes required by relaxing \emph{freshness} requirements (how up to date the state is after crash recovery). One popular approach is called group commit~\cite{helland1987group}. In this approach, log records are persisted in batches (or windows), such that the number of unpersisted records at any time is limited by a threshold, often called the group commit size.
By controlling the group commit size, the desired level of freshness can be achieved.

With PMEM, the performance characteristics of persistent writes have changed drastically. PMEM write latency and granularity is much lower than other PCIe connected storage devices, so the cost of persistence is lower too. This seems to imply that there is no merit in limiting write frequency with PMEM. However, we find that when adding replication overhead to overall performance characteristics, the overhead of persistence remains relatively high. This is because the cost of replication over RDMA is often an order of magnitude higher than local writes to PMEM. To reduce the cost of replication, we first explore if group commit can be a viable approach. We find that although it improves overall latency, it is not scalable; at high concurrency, group commit has significant overhead. This is because a shared counter or mutex is required to maintain the current window. Concurrent access to this counter/mutex across threads results in significant cache thrashing and reduces the effectiveness of group commit. This effect was not observed in prior work either because they did not allow concurrent access to the log, or because the IO latency overshadowed synchronization overhead. With fast PMEM hardware, the software and synchronization overhead constitutes a significant portion of overall latency, so the effects of synchronization are much more pronounced.

\begin{figure}[t]
\centering
\includegraphics[width=0.3\textwidth]{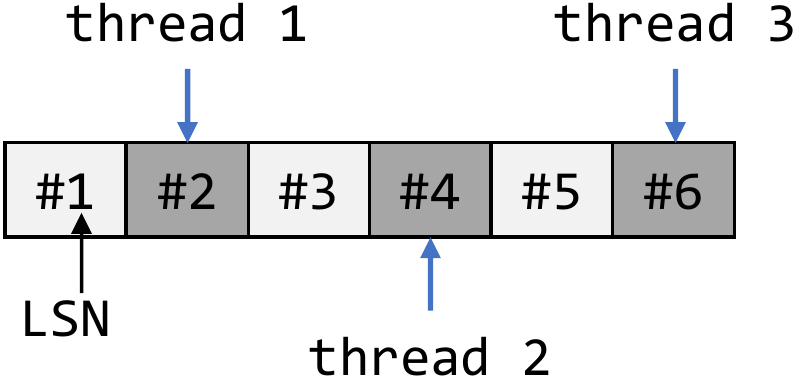}
\MyCaption{Example showing the worst case scenario for the frequency-based force policy with $T$=$3$ and $F$=$2$, where all threads are blocked in trying to force persistence. All 6 records have been completed and may be lost on crash.}
\label{fig:force-freq}
\end{figure}

To reduce the overhead of replication and persistence without adding synchronization overhead, we propose a novel frequency-based force policy. The key idea is as follows. Instead of maintaining a window of unpersisted records, we force records at intervals determined by a predefined `update frequency'. The crux of the idea lies in the fact that we can leverage the monotonicity of record LSNs to determine when to force data. For instance, given a frequency of $F$, data will be persisted every time a record with an LSN divisible by $F$ is forced. Each thread that completes a record with such an LSN becomes the `force leader' for the current batch of unpersisted records and is responsible for persisting them, 
which distributes this effort over all active threads.
In this manner, we do not need a shared counter or mutex to maintain the window, but instead piggy back on existing synchronization to generate monotonic LSNs (in reserve) as consecutive integers.

In our design, threads can write and force records to the log concurrently. Though the writes never overlap each other in the log space, they do race each other in time, and as a result, holes can be left in the log if a crash occurs after a thread writing further ahead in the log has completed its write while other threads haven't yet filled in the records leading up to it. In order to provide total ordering in Arcadia, a force operation cannot return success until all records leading up to the forced record have been persisted and marked complete, thus a force later in the log must wait for any earlier force operations.

In the worst case, all threads can get blocked doing a force. This can happen if, for instance, the first thread doing a force is context switched, or if a record with an earlier LSN that is not a multiple of $F$ has not yet completed. When threads force with the configured frequency, they are each separated by $F$ records. This gives us the theoretical upper bound of the number of completed records that can be lost as $F \times T$, where $T$ is the maximum number of threads or concurrent forces allowed. Figure~\ref{fig:force-freq} shows an example of the worst case scenario with $T$=$3$ and $F$=$2$. All three threads are blocked in this scenario. It is easy to see that the maximum number of records that may be lost on crash\footnote{We call this the vulnerability window.} can be correctly calculated using the formula described here.

\begin{figure*}[t]
\centering
\mbox{
    \subfigure[Single-Thread Latency]
    {
    	\label{fig:sc-lat}
    	\includegraphics[width=0.23\textwidth]{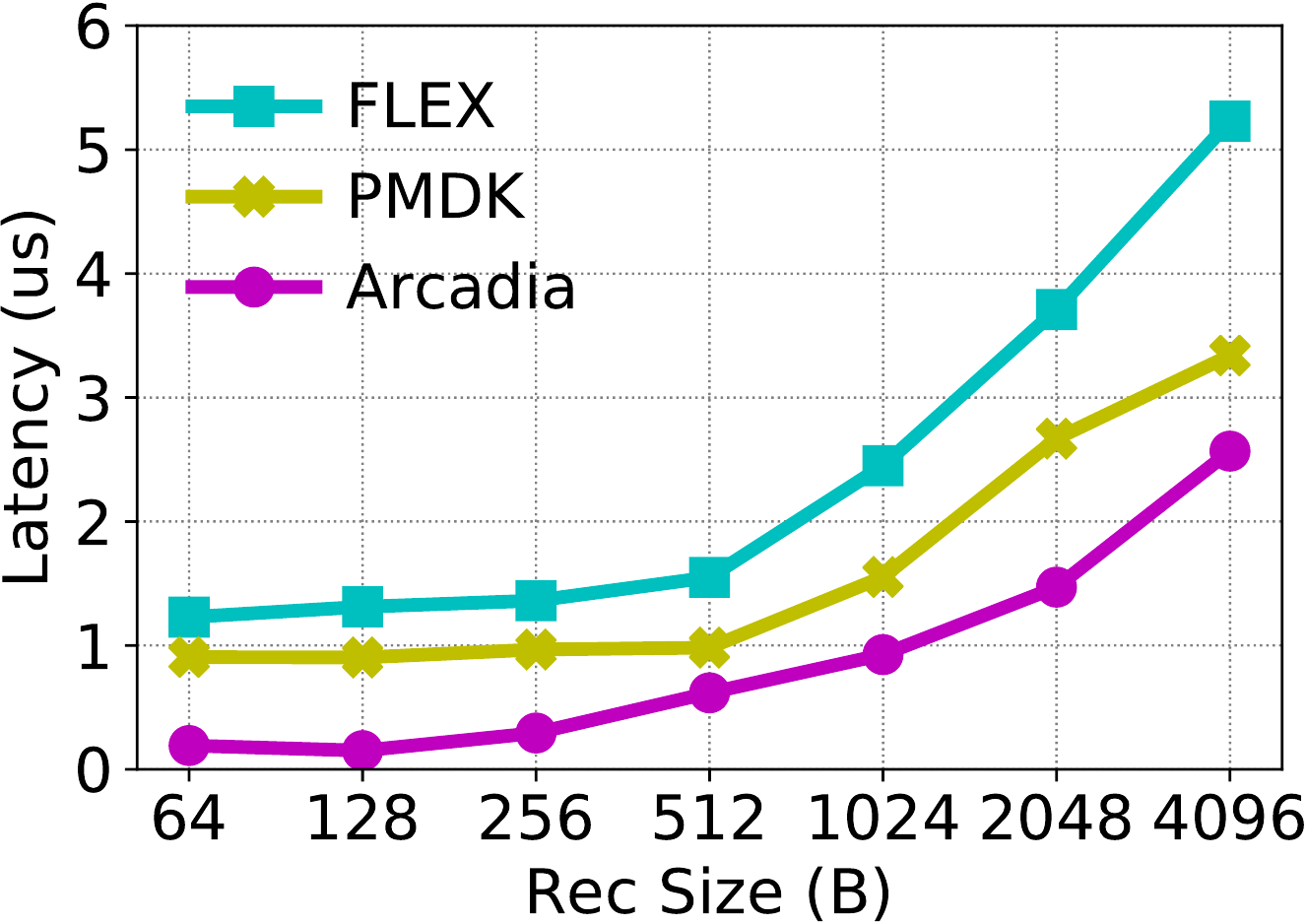}
    }
    \subfigure[Latency Breakdown]
    {
    	\label{fig:sc-lat-bd}
    	\includegraphics[width=0.23\textwidth]{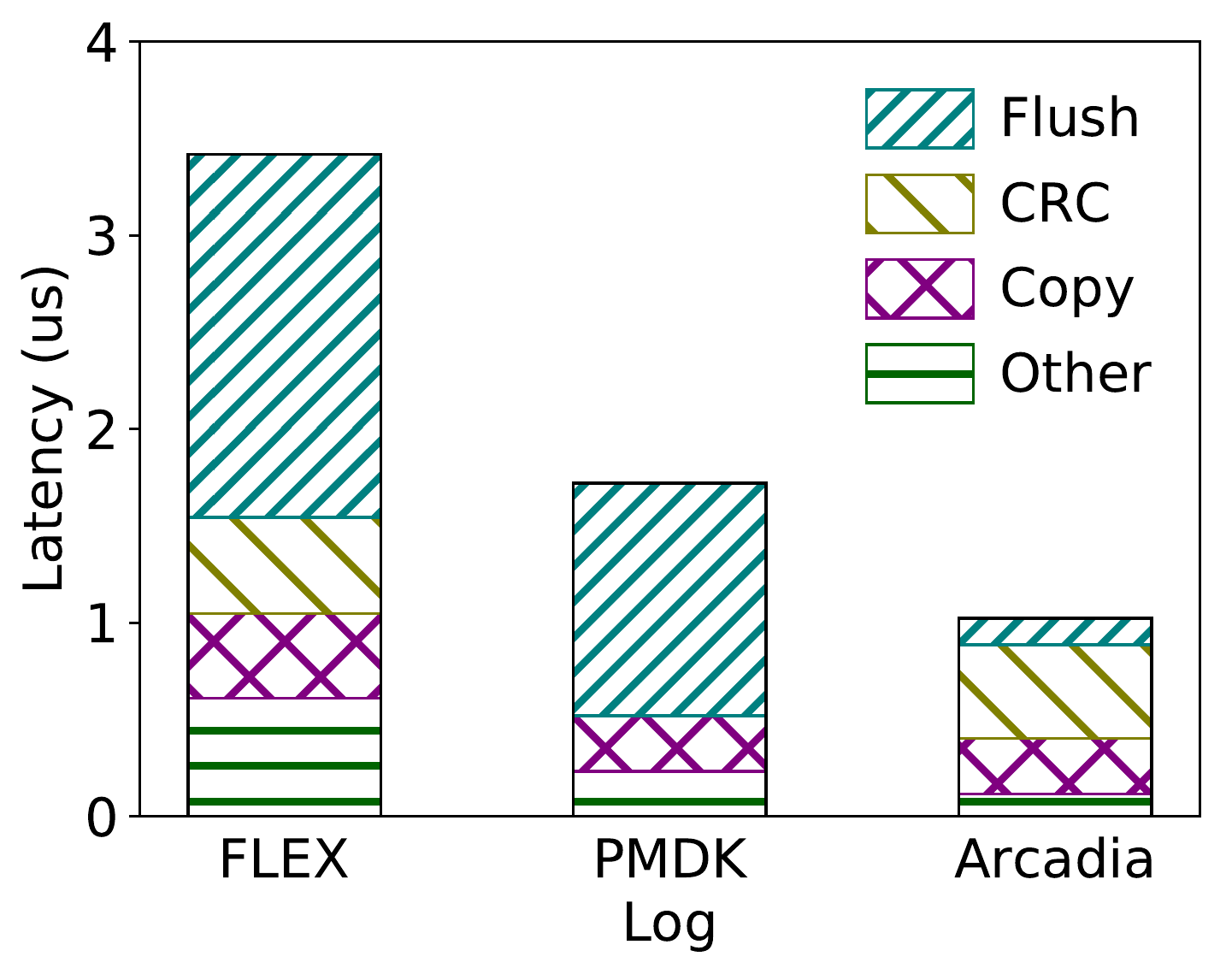}
    }
    \subfigure[Multi-Thread Throughput]
    {
    	\label{fig:mc-throughput}
    	\includegraphics[width=0.23\textwidth]{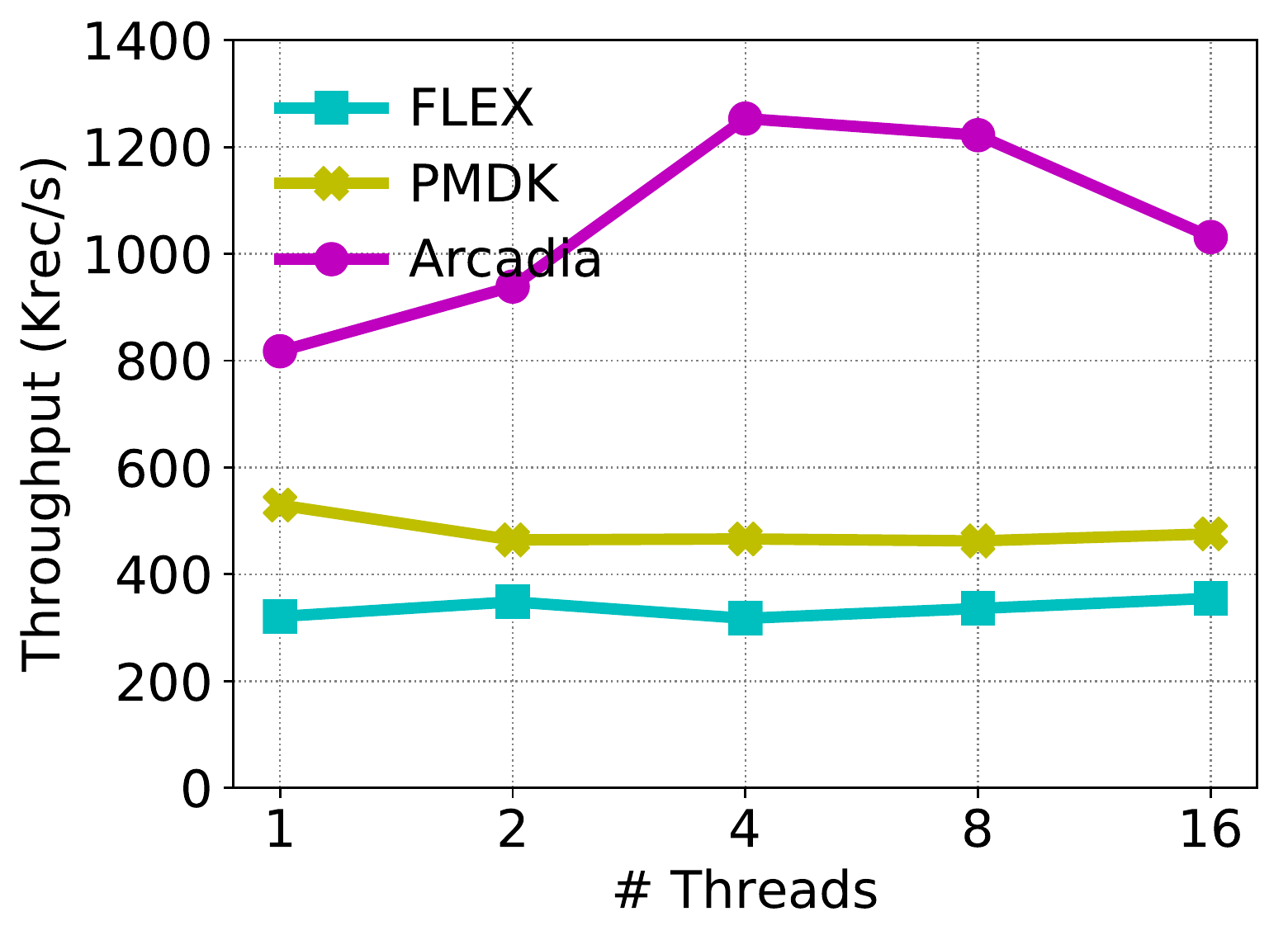}
    }
    \subfigure[Multi-Tenant Throughput]
    {
    	\label{fig:mt-throughput}
    	\includegraphics[width=0.23\textwidth]{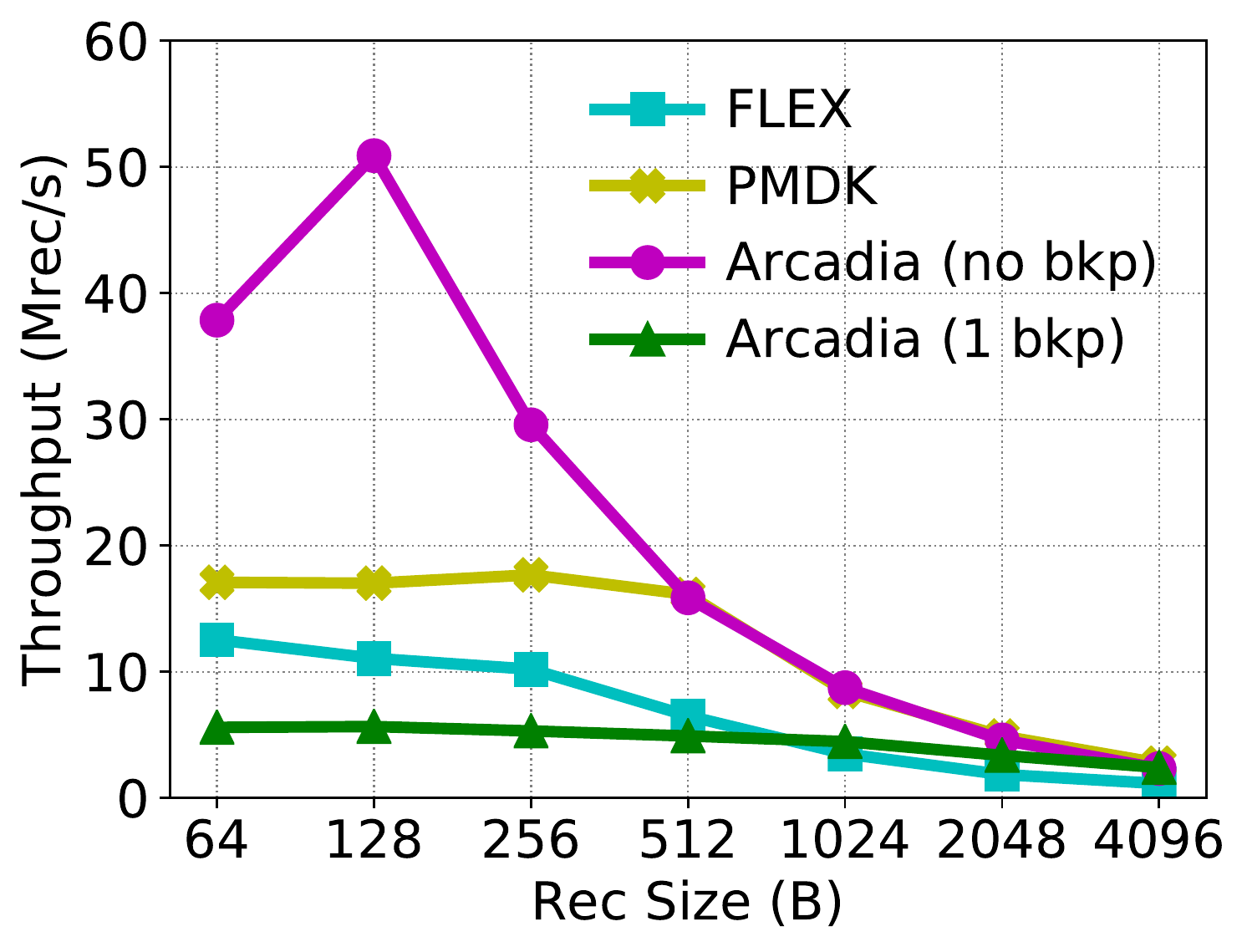}
    }
}
\MyCaption{Microbenchmark Comparison with FLEX and PMDK}
\label{fig:micro}
\end{figure*}

\noindent \textbf{Use Cases.} There are two use cases for this policy. The first is where explicit guarantees of persistence are not required for ensuring consistency (e.g., when recording logical operations in the log for a volatile key-value store). In this case, the frequency-based force policy can be used for all log updates to reduce the overhead of persisting records. A few of the recent updates may be lost on crash (the policy provides a bounded guarantee for how many updates can be lost), but consistency is not compromised. The second use case is where explicit guarantees of persistence are required for ensuring consistency (e.g., when recording state updates in the log for a database). Even in this case, the frequency-based force policy can be used for all log updates to reduce the overhead of persistence (by calling force with freq=$F$ for every record). When users require an explicit guarantee of record persistence (e.g., during the commit of a transaction), they may either issue a synchronous force (a force with freq=$1$) \emph{or} check if the difference between the LSN of a newly allocated record and the record they want forced is greater than the theoretical upper bound of the size of the vulnerability window, $F\times T$.

%% file: Texts/evaluation.tex
\MySection{Experimental Analysis}

In this section, we present the evaluation of Arcadia. We conduct a comprehensive analysis of the impact of the novel aspects of our design. We also compare performance and resilience of Arcadia with state-of-the-art PMEM-optimized logs, FLEX~\cite{xu2019finding}, PMDK's libpmemlog~\cite{pmdk} (referred to as simply PMDK in the evaluation), and Query Fresh~\cite{wang2017query}. We were unable to compare with Tailwind~\cite{taleb2018tailwind} because it requires special hardware support and its source code is unavailable.

\MySubsection{Experimental Testbed}

Our experimental testbed consists of two Linux (4.18) nodes, each equipped with two Cascade Lake CPUs (5218@2.30GHz), 192GB DRAM (2 x 6 x 16GB DDR4 DIMMs), and 1.5TB PMEM (2 x 6 x 128GB DCPMMs) configured in App Direct mode. Each CPU has 16 physical cores (with hyperthreading enabled) and 22MB of L3 cache (LLC). All available PMEM is formatted as two \code{ext4} DAX filesystems, each utilizing the memory of a single CPU socket as a single interleaved namespace. The two nodes are connected using 100Gbps EDR InfiniBand. One node is used as the primary while the other is used as a backup.

All code was compiled using GCC 8.3.1. PMDK~\cite{pmdk} 1.8 was used across all evaluations to keep comparisons fair. Hardware counters were obtained using a combination of Intel VTune Profiler~\cite{vtune}, Intel PCM~\cite{pcm}, and Linux perf utility. DCPMM hardware counters were collected using the \code{ipmwatch} utility, a part of the VTune Profiler.

\MySubsection{Microbenchmark Evaluation}

We first present a microbenchmark-level comparison of Arcadia with FLEX and PMDK. Since FLEX and PMDK do not support replication, we compare them with Arcadia deployed in local mode.

\noindent \textbf{Latency:} We evaluate the latency of log writes with a single thread while varying the record size. Figure~\ref{fig:sc-lat} shows this evaluation. We find that Arcadia has the lowest latency out of all three designs, up to 6x faster than PMDK and 8x faster than FLEX. One of the main reasons for this trend is that Arcadia does not maintain the pointer to the tail (most recently added record) of the log in PMEM (superline) and thus does not need to update it on every record write. FLEX performs the worst because it has high software overhead and it appends the record header and payload in separate operations. To confirm the reasons for these observations, we analyzed the breakdown of log writes for a 1KB record in Figure~\ref{fig:sc-lat-bd}. We can clearly observe that flushing data to PMEM takes much more time in FLEX and PMDK. This is because they both update the log tail for each write. The flush time is especially high because it also includes the additional store fence required to wait for data being copied into PMEM before the tail update. By avoiding a tail update, Arcadia shows much lower latency despite computing checksums.

\begin{figure*}[t]
\centering
\mbox{
    \subfigure[Log Write Latency]
    {
    	\label{fig:flush-policy-lat}
    	\includegraphics[width=0.23\textwidth]{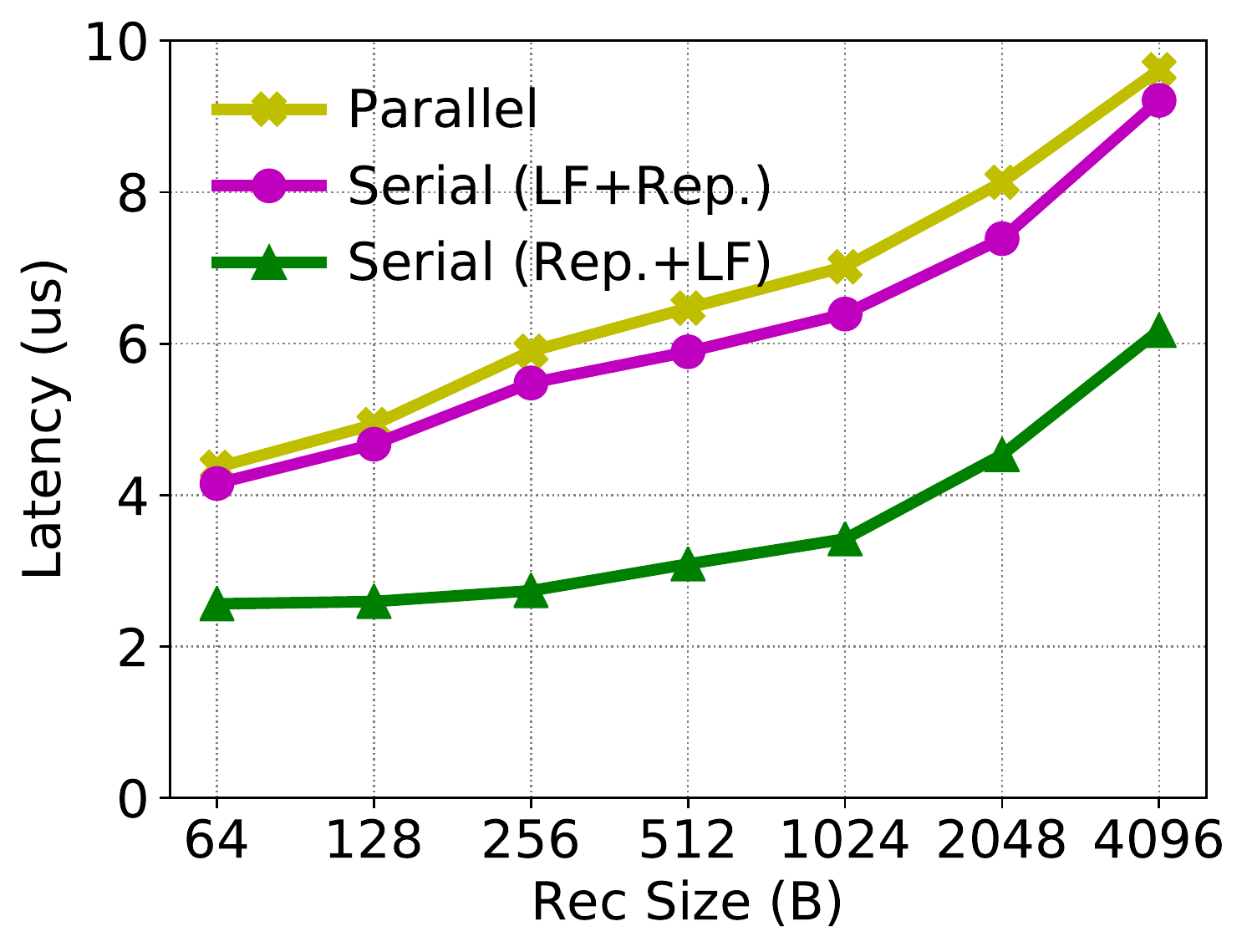}
    }
    \subfigure[Copy+Force Latency]
    {
    	\label{fig:flush-lat}
    	\includegraphics[width=0.22\textwidth]{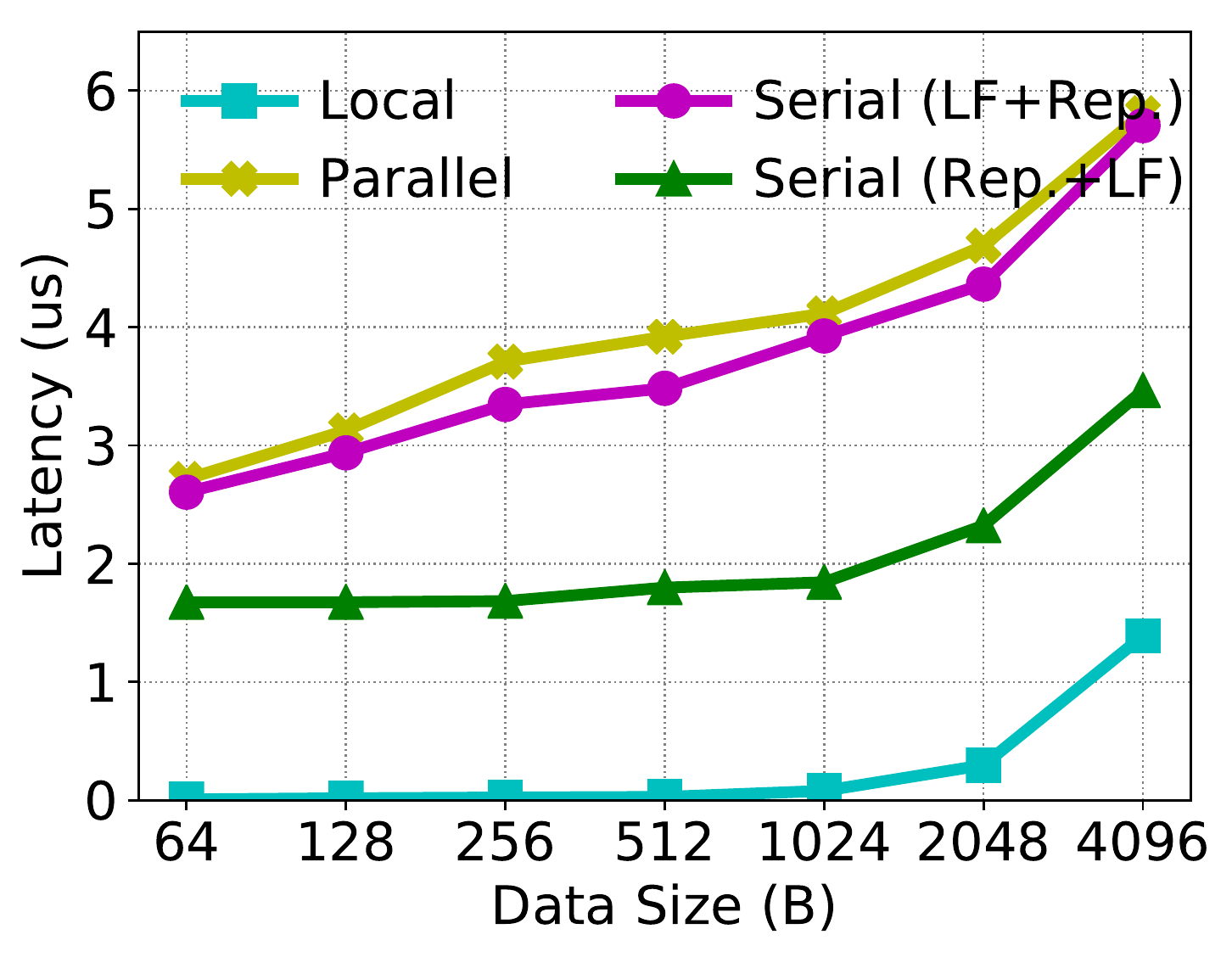}
    }
    \subfigure[LLC Misses]
    {
    	\label{fig:flush-llc-miss-rate}
    	\includegraphics[width=0.24\textwidth]{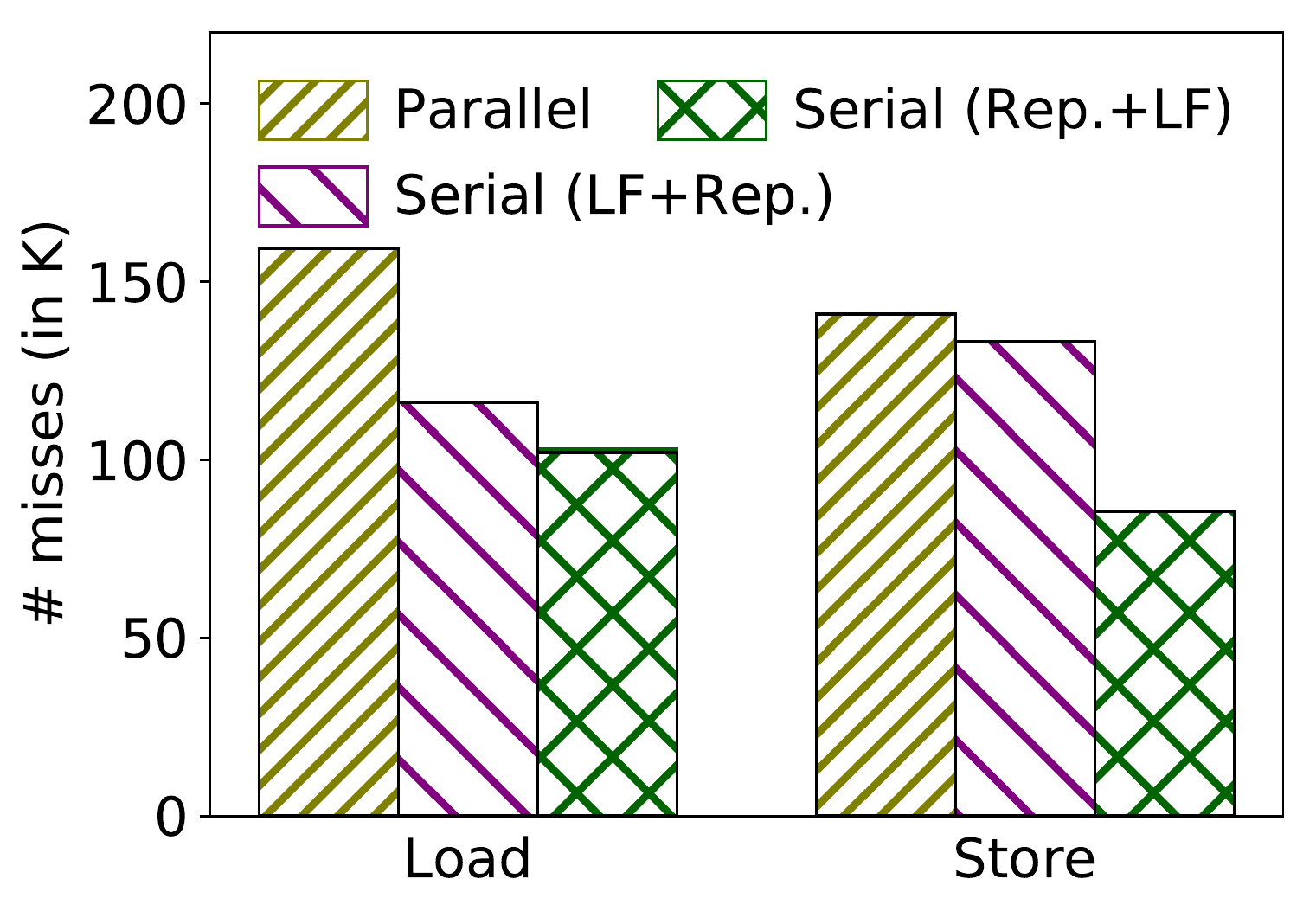}
    }
    \subfigure[Multi-Backup Throughput]
    {
    	\label{fig:mb-throughput}
    	\includegraphics[width=0.23\textwidth]{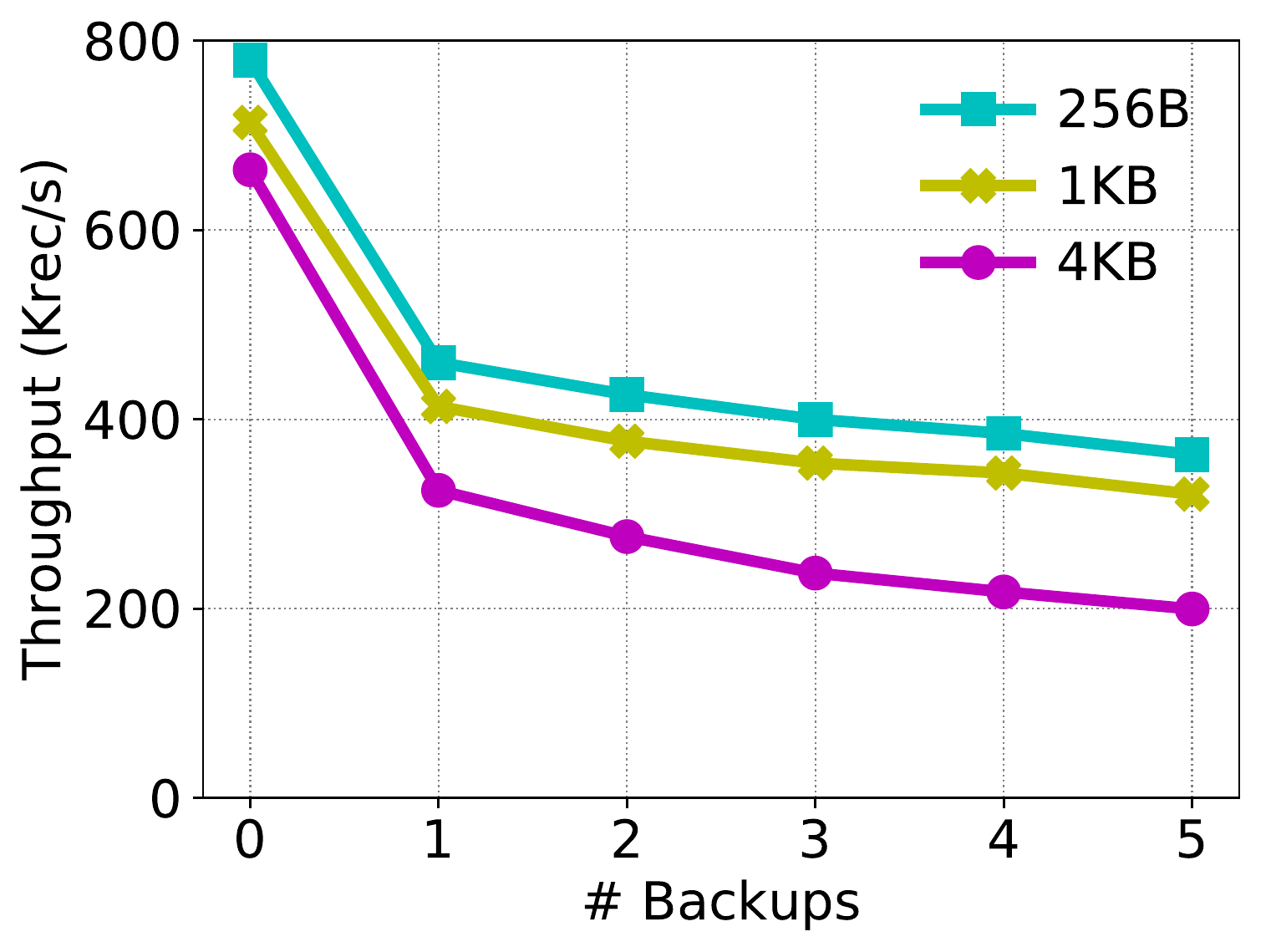}
    }
}
\MyCaption{Replication Overhead Analysis}
\label{fig:replication}
\end{figure*}

\noindent \textbf{Throughput:} Figure~\ref{fig:mc-throughput} presents throughput curves in a multi-threaded setting. We measure the overall log throughput while increasing the number of threads concurrently adding records to the log. Both FLEX and PMDK have flat curves because they fully isolate writers using coarse-grained locks and do not provide any concurrency. Arcadia only isolates steps that require serialization (reserve and force), and can provide some level of concurrency. Its throughput is maximum at 4 threads, but reduces slightly at higher concurrency. There are two reasons for this -- (1) synchronization overhead in reserve and force, and (2) poor bandwidth for highly concurrent accesses to PMEM. We also measure the aggregate log throughput in a multi-tenant setting. This is a common usage scenario where multiple tenants are sharing the same resources. Figure~\ref{fig:mt-throughput} shows the throughput for different record sizes with 16 concurrent single-threaded tenants, each writing to a separate log. Without replication, Arcadia performs the best in all cases. As the record size increases, throughput is bound by the PMEM bandwidth, and all log implementations converge to that limit. An observation we make is that Arcadia's throughput is lower for 64B records than 128B records. This is because of the write amplification effects in DCPMM for small-sized writes. With replication, Arcadia has lower throughput than FLEX and PMDK for small record sizes. In these cases, replication overhead bounds overall throughput. However, for large record sizes (> 2KB), throughput is indistinguishable from the local mode.

\MySubsection{Replication Overhead Analysis}

To understand the performance characteristics of replication with PMEM and RDMA, we conduct an experiment to measure the replication overhead with Arcadia deployed in local+remote mode. To highlight performance impact, we measured three methods of write flush ordering: (1) parallel, in which the local cache flush and RDMA replication are done in parallel, (2) serial with the local flush performed first followed by the remote flush (LF+Rep.), and (3) serial with the remote flush done first followed by the local flush (Rep.+LF). Figure~\ref{fig:flush-policy-lat} analyzes the log write latency across record sizes for the three methods. Surprisingly, we find that the flush order has significant impact on overall latency. Contrary to expectation, parallel has the worst latency, while serial with remote flush first has the best. To understand why, we analyzed the overhead of just the copy and force operations of log writes (shown in Figure~\ref{fig:flush-lat}) as well as the LLC miss counts of the three methods (shown in Figure~\ref{fig:flush-llc-miss-rate}). The copy+force overhead curves mirror the log write latency curves, confirming that the local flush and fence operations are responsible for higher RDMA latency.  Local flush is an order of magnitude faster than remote flush; therefore, when using the parallel method, the local cache flush invalidates data from the LLC.\footnote{At the time of writing this paper, cache write-back instructions had not been implemented in Intel processors.} So, RDMA writes need to read data back from PMEM instead of reading it directly from the LLC. This additional read is manifested as additional LLC misses and is confirmed by Figure~\ref{fig:flush-llc-miss-rate}. The serial method with local flush first suffers from the same problem, and shows only marginally better performance compared to the parallel method. We suspect that this is because concurrent reads and writes to PMEM in the parallel method conflict and slightly reduce performance. The serial method with remote flushed first (Rep.+LF) has the best performance because RDMA writes can go data directly from the LLC --- the LLC misses graph confirms this conclusion.

Next, we analyze the impact of number of log backups on throughput. Our testbed has only two nodes, so we conduct this experiment on another cluster with six Skylake nodes and EDR InfiniBand\footnote{This cluster does not have real PMEM, so we emulate PMEM with DRAM. This is also why we do not use this cluster for all other tests.}. Figure~\ref{fig:mb-throughput} shows the impact of number of backups on throughput for different record sizes. A common observation across record sizes is that adding backups results in a significant drop in throughput. This is expected because the overhead of replication is significant. However, adding additional backups after the first one has little additional impact because the asynchronous nature of RDMA enables data to be sent to multiple backups in parallel.  This is a significant result because it shows that whenever replication is enabled, throughput is not significantly impacted by the number of backups. Consequently, it's not necessary to compromise fault-tolerance for performance when replication is enabled. These results also show that in cases where PMEM is not available on all nodes, even nodes without PMEM can use Arcadia efficiently by having only remote log replicas.

\MySubsection{Recovery Evaluation}

\begin{figure}[t]
\centering
\mbox{
    \subfigure[Recovery Latency]
    {
    	\label{fig:recovery-lat}
    	\includegraphics[width=0.23\textwidth]{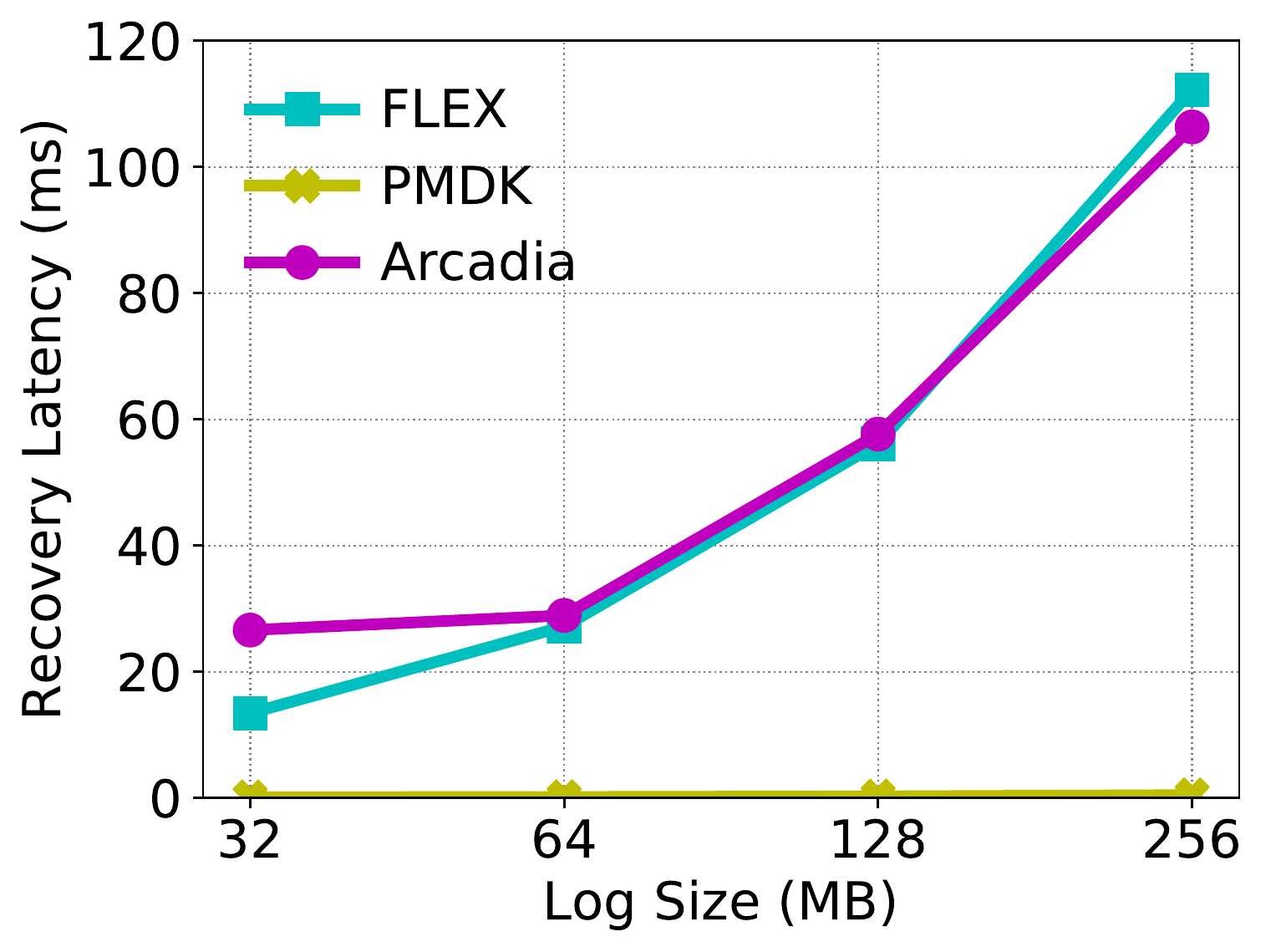}
    }
    \subfigure[Arcadia Recovery Latency]
    {
    	\label{fig:Arcadia-recovery-lat}
    	\includegraphics[width=0.23\textwidth]{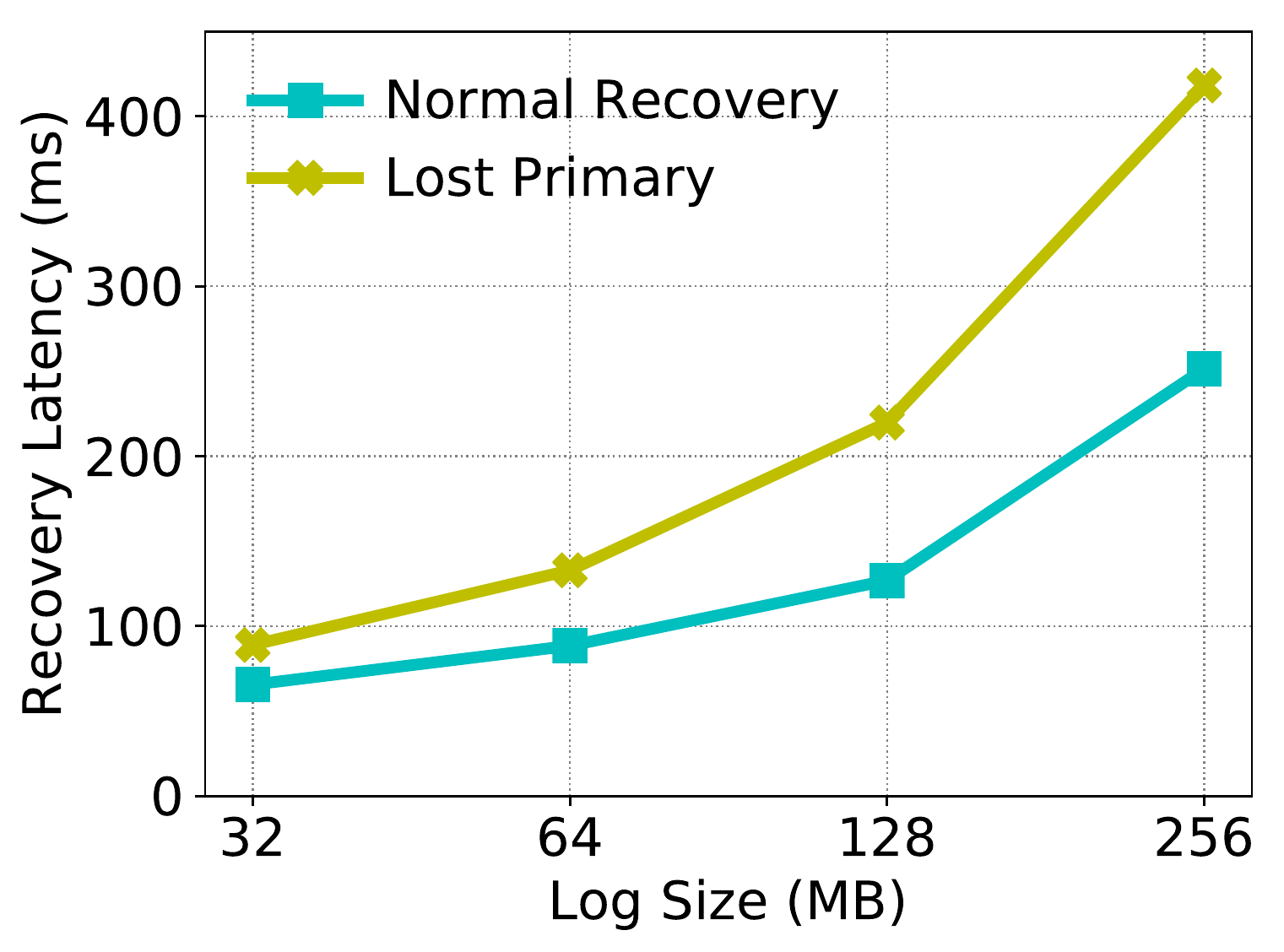}
    }
}
\MyCaption{Recovery Evaluation}
\label{fig:recovery}
\end{figure}

\begin{figure*}[t]
\centering
\mbox{
    \subfigure[Impact on Throughput]
    {
    	\label{fig:force-policy-throughput}
    	\includegraphics[width=0.24\textwidth]{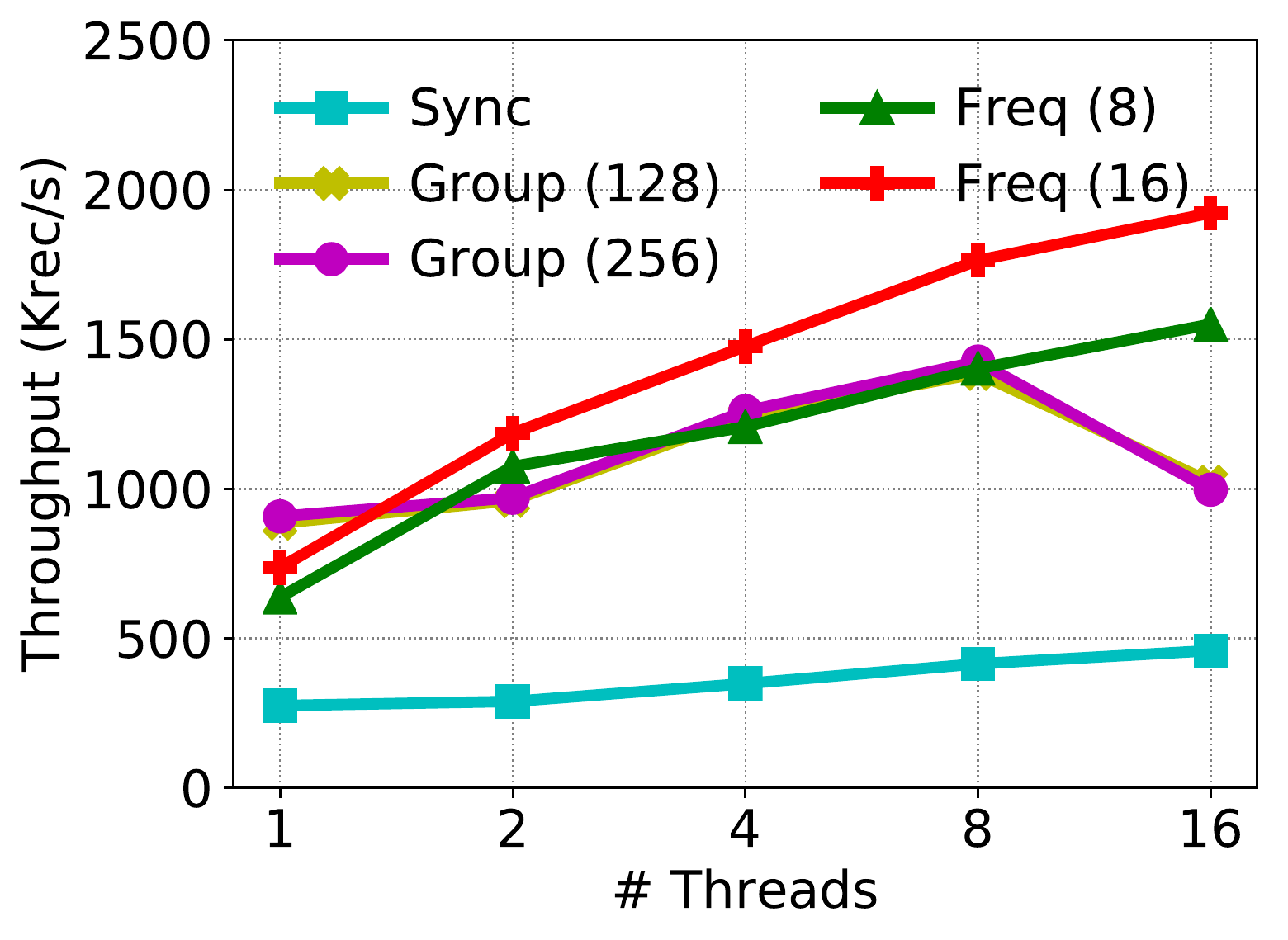}
    }
    \subfigure[Impact on L1d Misses]
    {
    	\label{fig:force-policy-l1-miss-rate}
    	\includegraphics[width=0.22\textwidth]{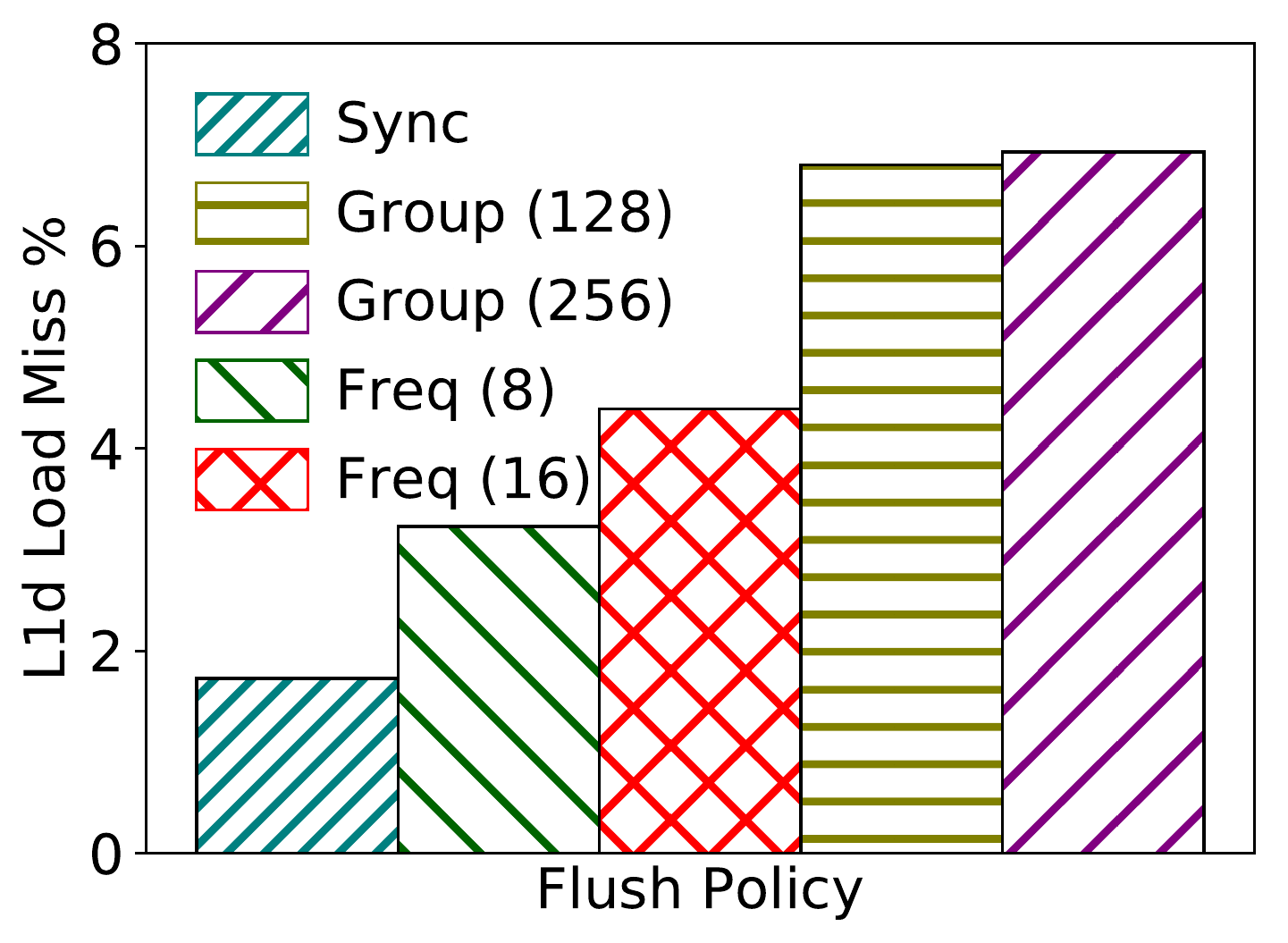}
    }
    \subfigure[Vulnerability Window (Freq=8)]
    {
    	\label{fig:window-freq-8}
    	\includegraphics[width=0.23\textwidth]{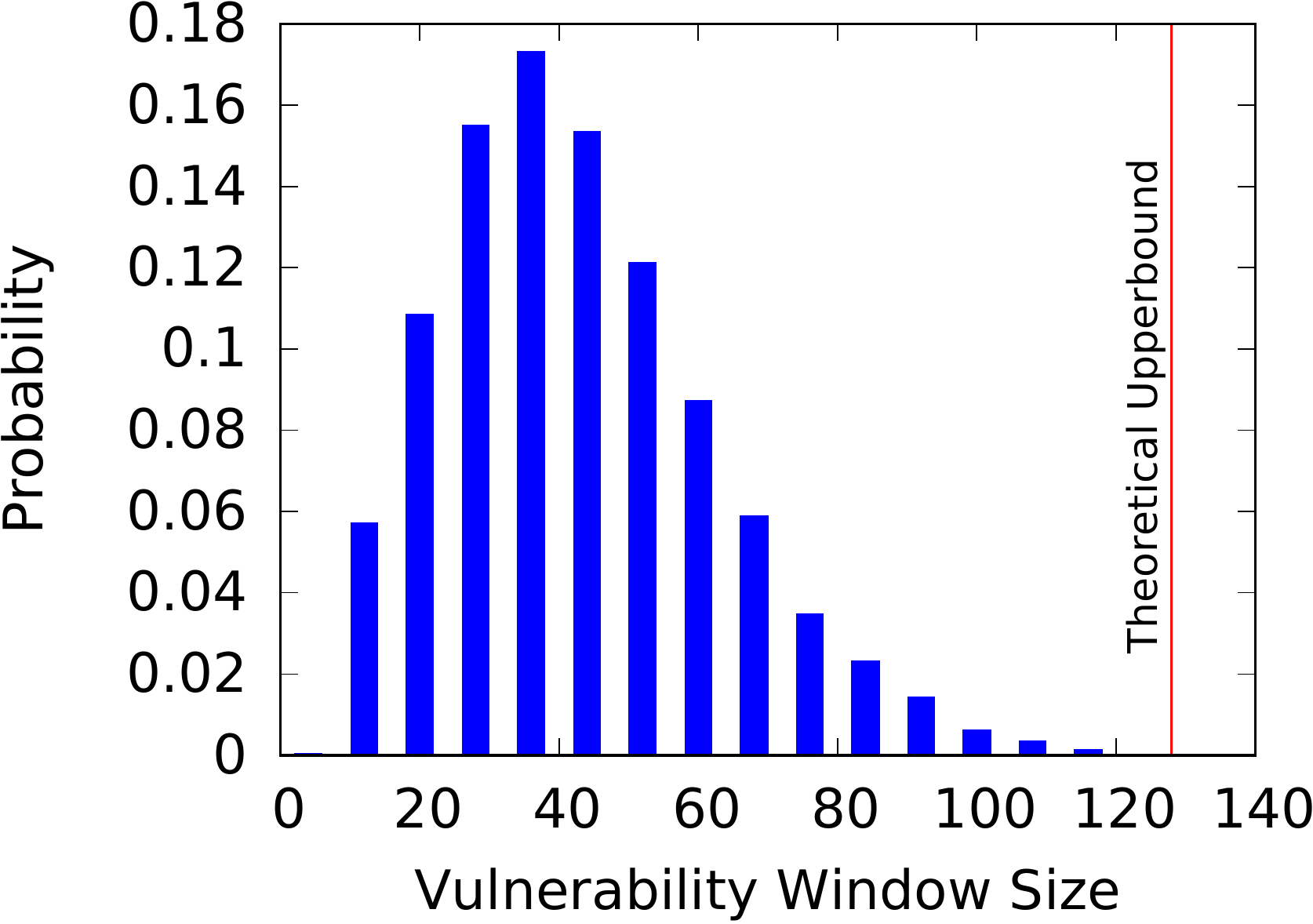}
    }
    \subfigure[Vulnerability Window (Freq=16)]
    {
    	\label{fig:window-freq-16}
    	\includegraphics[width=0.23\textwidth]{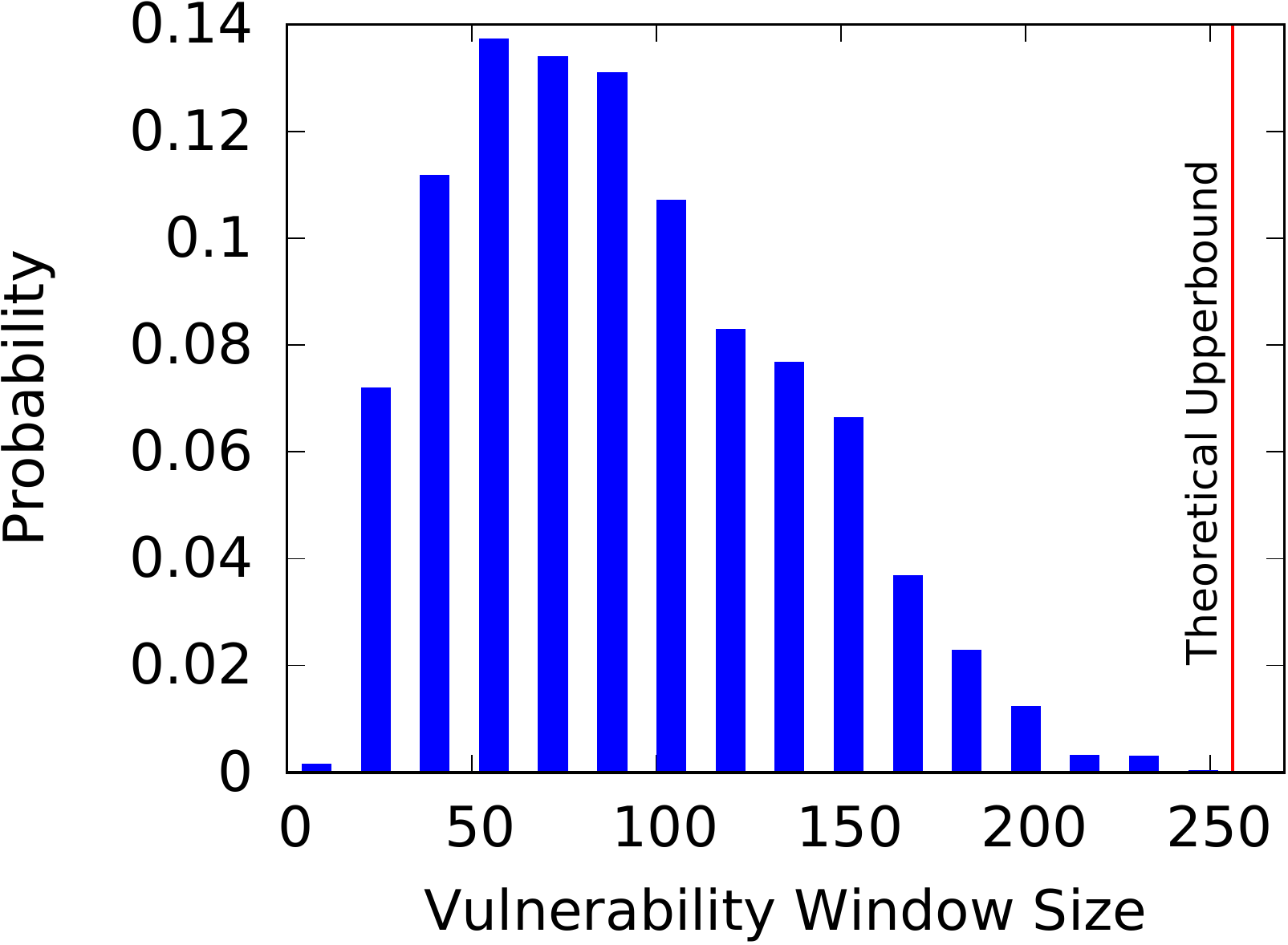}
    }
}
\MyCaption{Force Policy Analysis}
\label{fig:force-policy}
\end{figure*}

To evaluate the recovery performance of Arcadia, we measured the total time taken to recover log state and iterate over all valid records on recovery, calling a null recovery function for each record. Figure~\ref{fig:recovery-lat} compares Arcadia's recovery latency in local mode with FLEX and PMDK. Both Arcadia and FLEX rely on checksums to verify record integrity, so they have similar performance. This is because the time taken to verify checksums dominates recovery latency, which is also why latency increases linearly with log size. PMDK does not use checksums, so its recovery procedure only consists of calling null functions for each record. This is why it is able to recover so fast. However, by not relying on any integrity checks, it is unable to handle undetected media errors and may redo or undo updates using corrupted data. Figure~\ref{fig:Arcadia-recovery-lat} compares Arcadia's recovery latency with replication enabled in normal recovery and when the primary log copy fails or is lost. These represent the best and worst case situations for recovery, respectively. We observe that when the primary copy is lost, the latency is higher. In this scenario, Arcadia needs to recover the primary copy using data in the backup which increases recovery time. However, this process uses the fast one-sided RDMA reads for copying data which does not add significant overhead. Even for a 256MB log, recovery takes less than 500ms in the worst case; so, we conclude that recovery is fast enough for most practical scenarios.

\MySubsection{Force Policy Analysis}

We compare group commit and frequency-based force policies in this experiment. We compare log throughput for the two policies in Figure~\ref{fig:force-policy-throughput}. The group size or frequency is listed in parenthesis. We evaluate with two different values for each policy; these values have been chosen such that their theoretical vulnerability window sizes are comparable. So, group (128) is comparable to freq (8), while group (256) is comparable to freq (16). We also evaluate the sync policy in which each log write is synchronously forced. Results demonstrate the group commit has significant overhead at high concurrency and its throughput drops by 30\% at 16 threads. Concurrent access to the window size variable is behind this drop. On the other hand, the frequency-based policy is much more scalable because it avoids this synchronization overhead. To confirm that synchronization overhead is the reason for poor performance of group commit, we analyze the L1d miss rate for all policies (see Figure~\ref{fig:force-policy-l1-miss-rate}). We can clearly observe that group commit has much higher miss rates because of constant cache thrashing as a result of concurrent accesses to the shared counter.

We also measure the distribution of the vulnerability window size (from latest completed record to the most recently forced record) for the frequency-based policy. Unlike group commit, the vulnerability window size is not fixed because if a thread gets blocked doing a force, other threads can keep adding new records and increase the window size. Figures~\ref{fig:window-freq-8} and~\ref{fig:window-freq-16} show this distribution for frequencies of 8 and 16, respectively. Interestingly, we find that the probability distribution is skewed towards smaller sizes, far below the theoretical upper bound. This shows that on average, threads are unlikely to block on a force. Overall, we conclude that the frequency-based policy is not only more scalable, but in practice it provides better resilience than its theoretical limit.

\MySubsection{Arcadia Applications}

\begin{figure}[t]
\centering
\mbox{
    \subfigure[Latency]
    {
    	\label{fig:rocksdb-latency}
    	\includegraphics[width=0.23\textwidth]{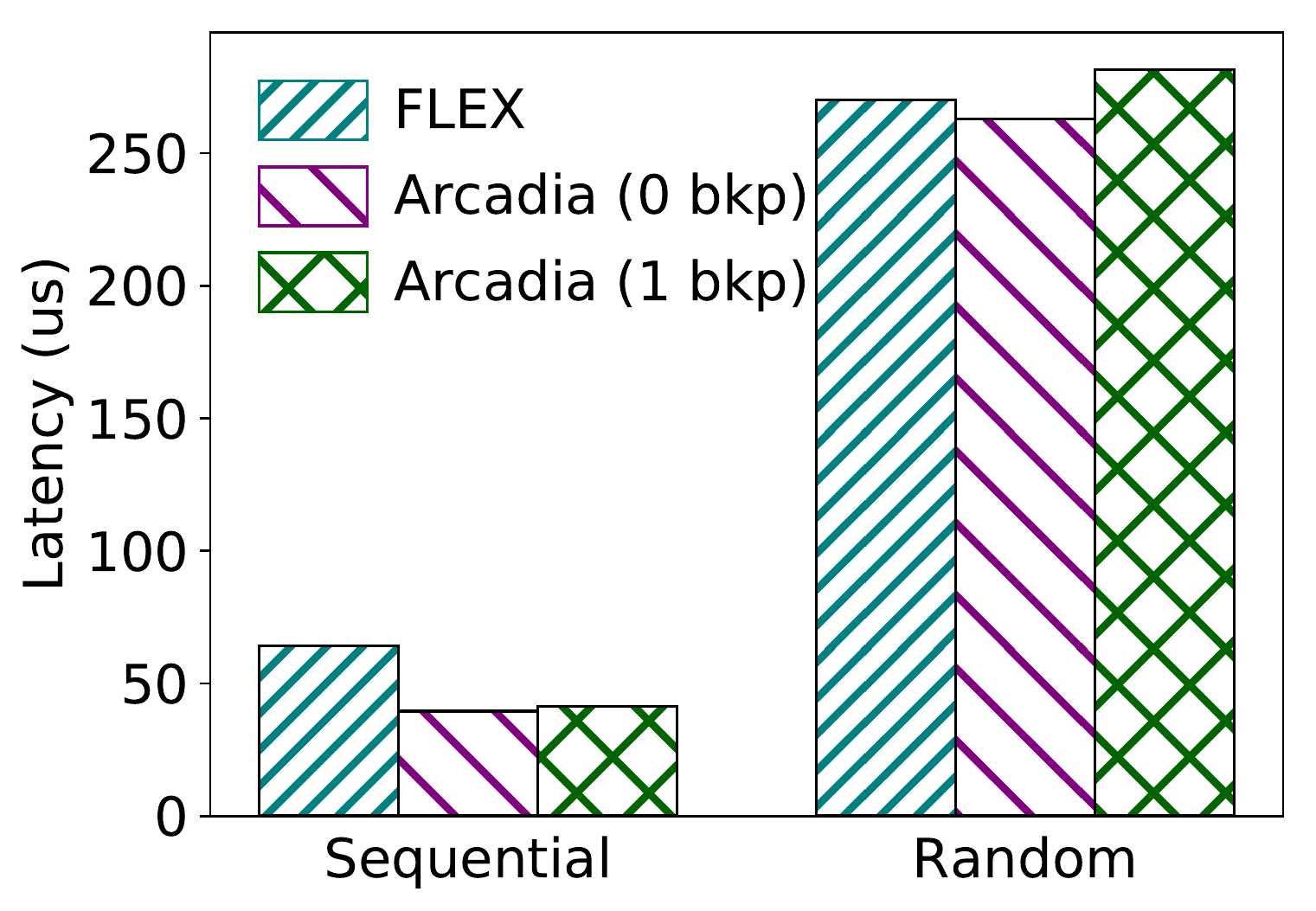}
    }
    \subfigure[Throughput]
    {
    	\label{fig:rocksdb-throughput}
    	\includegraphics[width=0.23\textwidth]{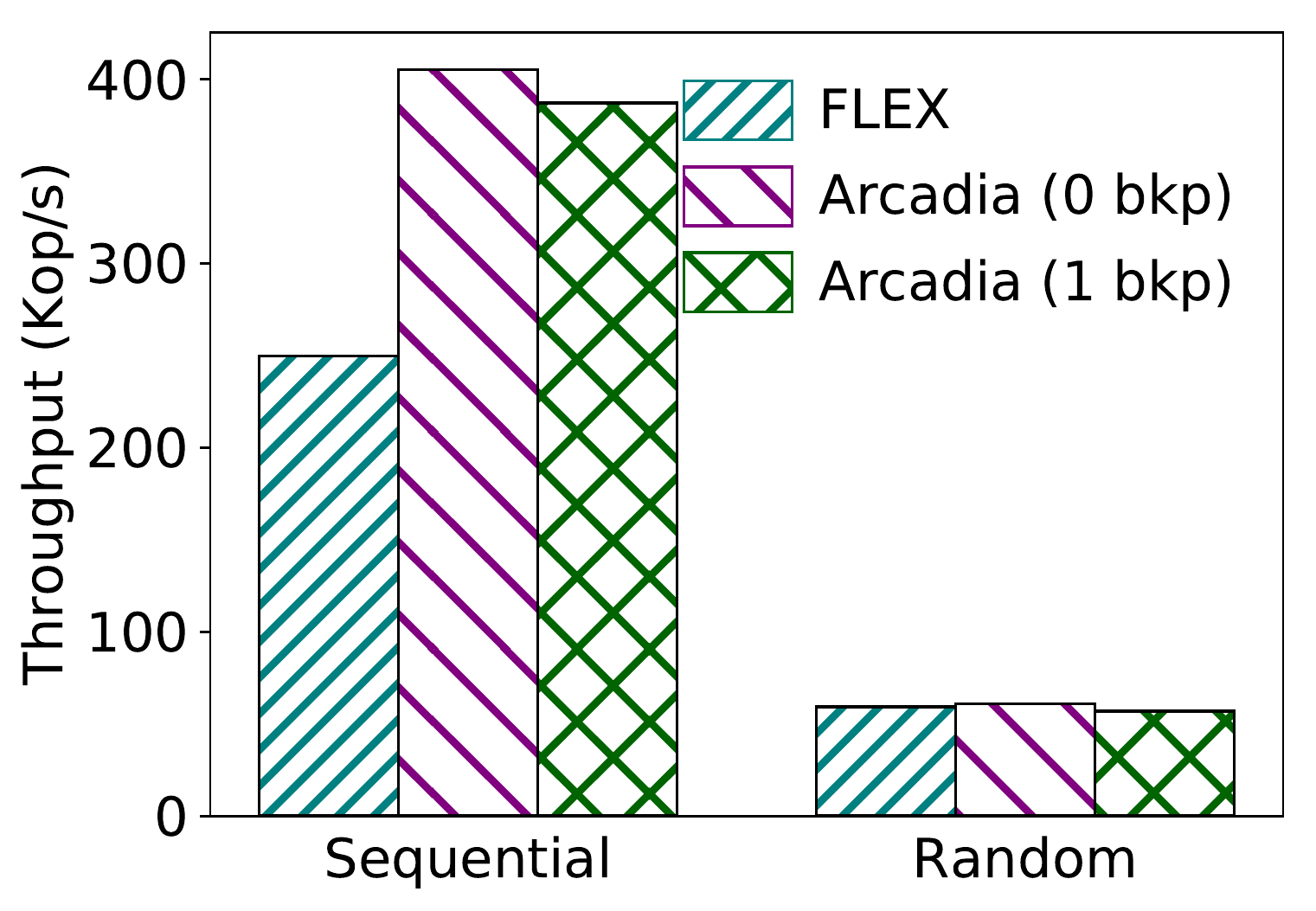}
    }
}
\MyCaption{Comparison with FLEX using RocksDB}
\label{fig:rocksdb}
\end{figure}	

\noindent \textbf{RocksDB.} To demonstrate the practical benefits of Arcadia, we integrate it with RocksDB (adding/changing only \textasciitilde 200 LoC), a popular key-value database used at Facebook, by swapping out its write-ahead-log with Arcadia and use the fine-grained interface for adding log records. We evaluate this version in both local and local+remote modes and compare it with the FLEX integration of RocksDB~\cite{pmem-rocksdb}. Figure~\ref{fig:rocksdb} compares the latency and throughput for both sequential and random key-value puts at full-subscription (16 threads). By reducing the log append latency and using the fine-grained API to increase concurrency, Arcadia improves latency by up to 38\% and throughput by up to 62\% in local mode (0 bkp). In local+remote mode (1 bkp), Arcadia is faster than FLEX (which is local-only) for sequential puts and on-par for random puts, despite enabling replication. Replication overhead is much less than the overhead of the entire put operation. Hence, there is little difference in performance with and without replication. Finally, random puts are much slower than sequential puts, so latency is dominated by operations other than logging, and log performance differences are negligible.

\begin{figure}[t]
\centering
\mbox{
    \subfigure[Throughput]
    {
    	\label{fig:masstree-throughput}
    	\includegraphics[width=0.23\textwidth]{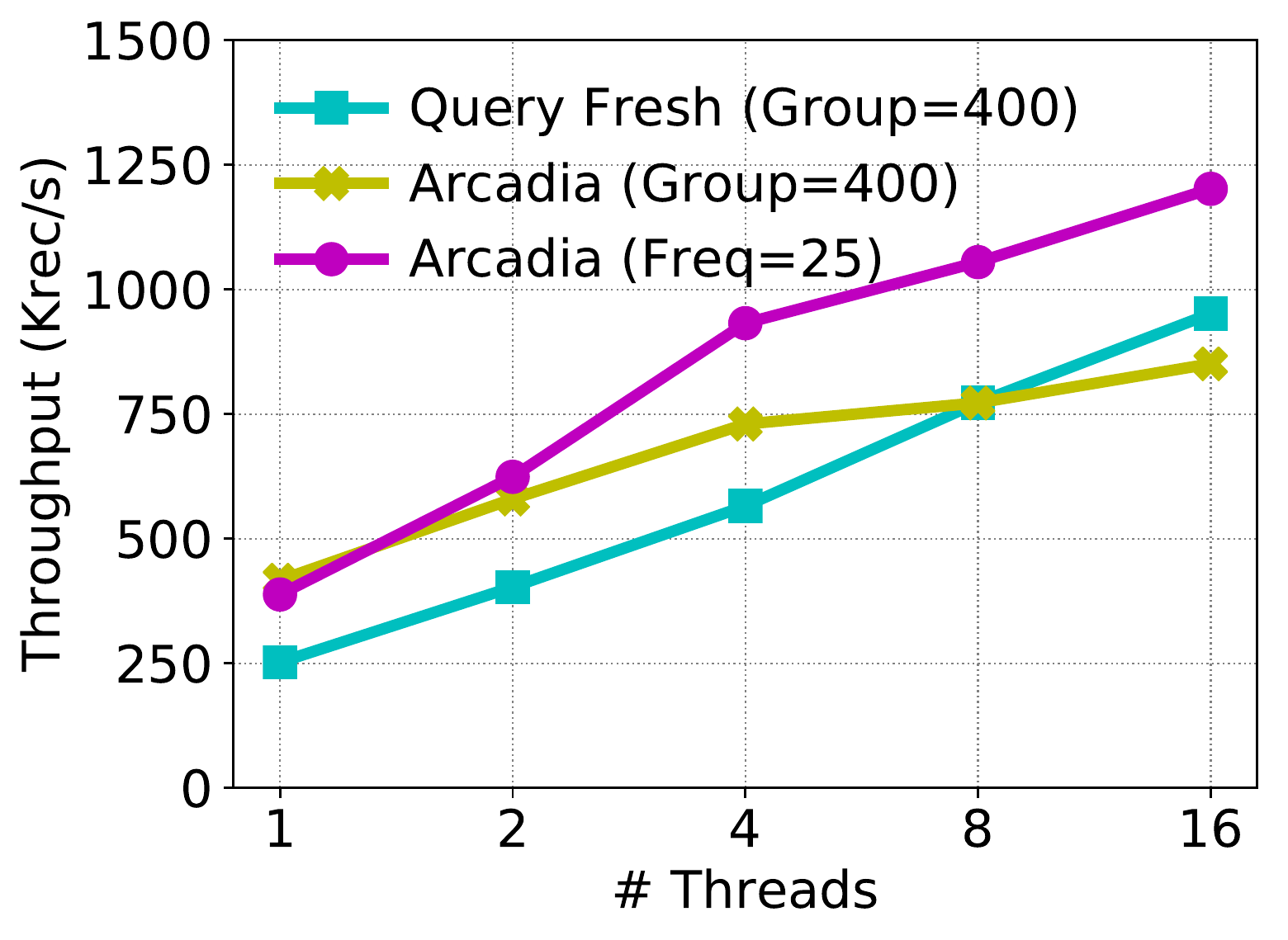}
    }
    \subfigure[Fault-Tolerance]
    {
    	\label{fig:masstree-ft}
    	\includegraphics[width=0.23\textwidth]{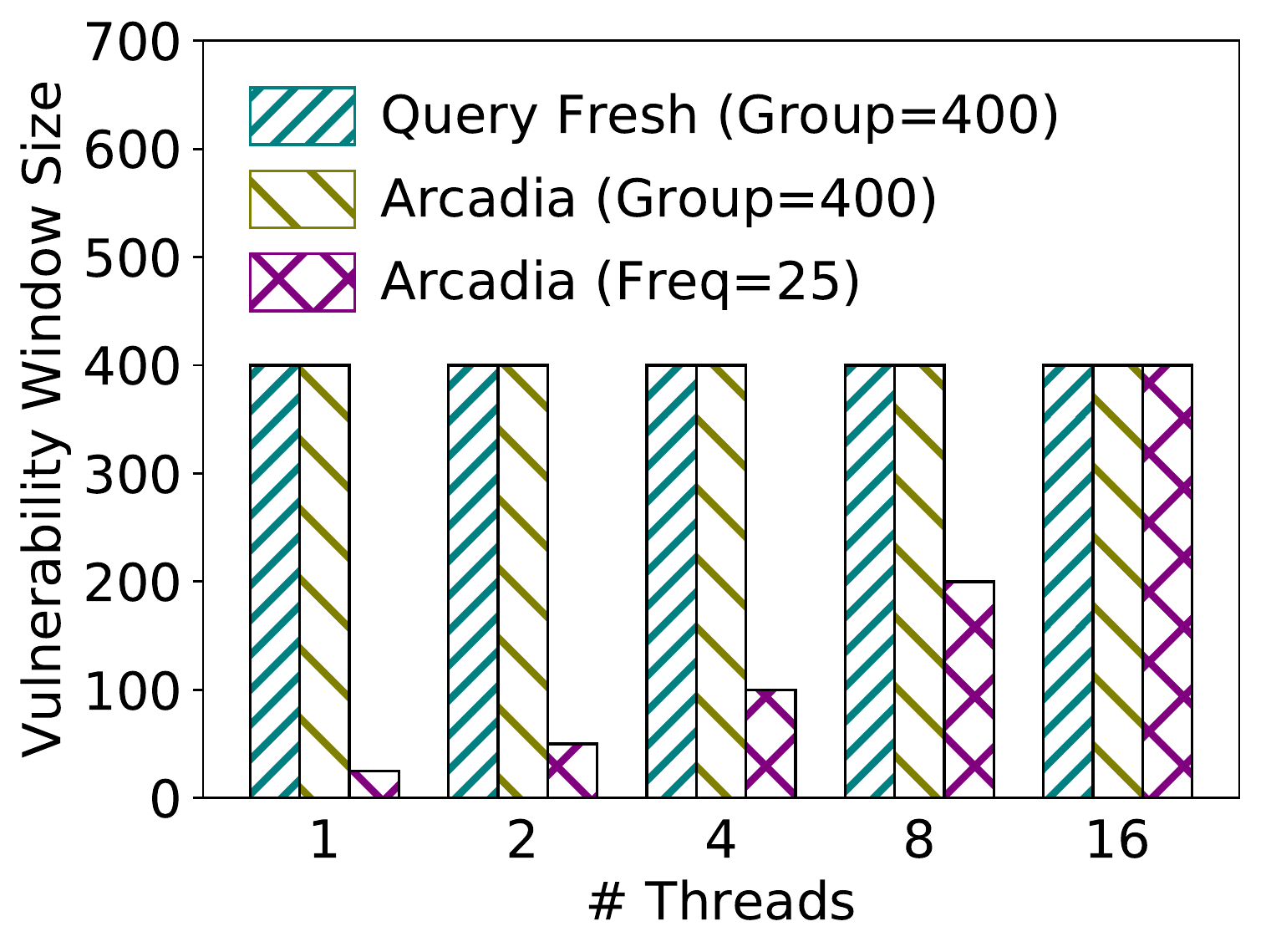}
    }
}
\MyCaption{Comparison with Query Fresh using Masstree}
\label{fig:queryfresh}
\end{figure}

\noindent \textbf{Masstree.} We integrate Arcadia with Masstree~\cite{mao2012cache} (adding/changing only \textasciitilde 100 LoC), another popular key-value database, to enable comparison with Query Fresh. This experiment shows the practical benefits of the frequency-based force policy. Figure~\ref{fig:masstree-throughput} compares the throughput of read-modify-writes and Figure~\ref{fig:masstree-ft} compares the theoretical vulnerability window size for Query Fresh and Arcadia (with both group commit and frequency-based force policies). From the results, we observe that Arcadia's group commit policy has high synchronization overhead at 8 and 16 threads which reduces performance bringing it close to Query Fresh's throughput. Query Fresh also uses group commit but it only enables limited log concurrency. Therefore, it delivers lower throughput than Arcadia but is also less impacted by the synchronization overheads of group commit. The frequency-based policy is able to deliver the best performance (up to 65\% faster than Query Fresh) and fault-tolerance by avoiding unnecessary synchronization and forcing records more frequently.

\MySubsection{Key Insights}

Our measurements demonstrate the performance benefits of our approach compared to PMDK, FLEX, and Query Fresh. The measurements confirm the benefits of specific innovations and design choices in Arcadia, such as avoiding the superline tail pointer, and using concurrency to limit the impact of checksums and replication.
They also highlight subtle factors such as the synchronization implied by group commit and the interaction between local flush and remote replication, giving confidence in the overall conclusions.

Arcadia's log interface crucially decouples steps requiring serialization from those allowing concurrency to maximize performance of parallel applications.
It also illustrates PMEM's byte-addressability to avoid unnecessary data copying.
The proposed frequency-based force policy gives flexibility in the freshness and performance trade-off, while allowing more concurrency than the traditional group commit method.

%% file: Texts/related.tex
\MySection{Related Work}
	
\noindent \textbf{Logging on PMEM.} A number of prior works~\cite{taleb2018tailwind,pmdk,xu2019finding,wang2017query,huang2014nvram,kim2016nvwal,arulraj2016write}, have leveraged PMEM performance for logging as a means to improve overall system performance. Some of these works (Query Fresh~\cite{wang2017query}, WBL~\cite{arulraj2016write}, NV-Logging~\cite{huang2014nvram}) propose PMEM-based logs that are tightly integrated with a particular system or logging algorithm. Some of them (PMDK~\cite{pmdk}, FLEX~\cite{xu2019finding}, NVWAL~\cite{kim2016nvwal}) propose unreplicated logs that cannot provide high availability. Others (e.g., Tailwind~\cite{taleb2018tailwind}) rely on special hardware support. Moreover, none of these are able to satisfy the robustness requirements of large-scale production systems because they do not provide resilience to all likely failure scenarios.

\noindent \textbf{Logging on Flash/Disk.} Corfu~\cite{balakrishnan2012corfu}, Raft-based log~\cite{ahn2019designing}, Apache Kafka~\cite{wang2015building}, DistributedLog~\cite{guo2017distributedlog}, GNR~\cite{deenadhayalan2011gpfs}, and JBD2~\cite{tweedie1998journaling} are relevant log implementations on traditional flash and disk storage that are deployed in production systems. These designs are considered to be reliable and robust. However, they were designed considering the performance characteristics, access methods, and reliability of flash/disk which are considerably different than those of PMEM. Naïvely applying them to PMEM is unlikely to yield good results.

\noindent \textbf{Combining RDMA and PMEM.} Octopus~\cite{lu2017octopus}, Forca~\cite{huang2018forca}, File-MR~\cite{yang2020filemr}, Orion~\cite{yang2019orion}, pDPM~\cite{tsai2020disaggregating}, AsymNVM~\cite{ma2020asymnvm}, NVFS~\cite{islam2016high}, Hotpot~\cite{shan2017distributed}, Mojim~\cite{zhang2015mojim}, and RDMP-KV~\cite{li2020rdmp} have worked on combining PMEM with RDMA for fast access to remote persistent storage. Each takes a different approach to remote persistence and atomicity that are not broadly applicable in other situations because they are either tied to a particular data format (key-value or file) \emph{or} the complications of RDMA + PMEM persistence are not fully solved. In contrast, in this paper, we present abstract primitives for ensuring remote persistence, atomicity, and integrity, that can be applied generically on commodity hardware.

%% file: Texts/conclusion.tex
\MySection{Conclusion}


We've described Arcadia, a generic log with easy to use interface, stored on PMEM, and replicated using RDMA.  Implementing fast, fault-tolerant systems is a challenging problem, and a good log implementation is a crucial abstraction for success.  Arcadia efficiently encapsulates the operations of PMEM and RDMA to manage atomicity, persistence, and replication. It takes a creative approach in its API to eliminate unnecessary synchronization to provide the concurrency necessary to get the full performance from PMEM.  We hope that wide adoption of this technology will significantly advance the frontier of fast and robust storage systems.

%% file: main.bbl
\begin{thebibliography}{10}

\bibitem{ahn2019designing}
Jung-Sang Ahn, Woon-Hak Kang, Kun Ren, Guogen Zhang, and Sami Ben-Romdhane.
\newblock {Designing an Efficient Replicated Log Store with Consensus
  Protocol}.
\newblock In {\em HotCloud}, 2019.

\bibitem{Akinaga2010ResistiveRA}
Hiroyuki Akinaga and Hisashi Shima.
\newblock {Resistive Random Access Memory (ReRAM) Based on Metal Oxides}.
\newblock {\em Proceedings of the IEEE}, 98:2237--2251, 2010.

\bibitem{apache-zookeeper}
{{Apache ZooKeeper}}.
\newblock \url{http://www.zookeeper.apache.org} (accessed Sep. 2021).

\bibitem{arulraj2016write}
Joy Arulraj, Matthew Perron, and Andrew Pavlo.
\newblock {Write-Behind Logging}.
\newblock {\em Proceedings of the VLDB Endowment}, 10(4):337--348, 2016.

\bibitem{roce2}
InfiniBand~Trade Association.
\newblock {Supplement to Infiniband Architecture Specification Volume 1,
  Release 1.2. 1: Annex A17: RDMA over Converged Ethernet (RoCEv2)}, 2014.

\bibitem{balakrishnan2012corfu}
Mahesh Balakrishnan, Dahlia Malkhi, Vijayan Prabhakaran, Ted Wobbler, Michael
  Wei, and John~D Davis.
\newblock {CORFU: A Shared Log Design for Flash Clusters}.
\newblock In {\em NSDI}, pages 1--14, 2012.

\bibitem{deenadhayalan2011gpfs}
Veera Deenadhayalan.
\newblock {GPFS Native RAID for 100,000-Disk Petascale Systems}.
\newblock In {\em LISA}, 2011.

\bibitem{snia1}
Chet Douglas.
\newblock {RDMA with PMEM}.
\newblock
  \url{http://www.snia.org/sites/default/files/SDC15_presentations/persistant_mem/ChetDouglas_RDMA_with_PM.pdf},
  2015.

\bibitem{guo2017distributedlog}
Sijie Guo, Robin Dhamankar, and Leigh Stewart.
\newblock {DistributedLog: A High Performance Replicated Log Service}.
\newblock In {\em ICDE}, pages 1183--1194, 2017.

\bibitem{hady2017platform}
Frank~T Hady, Annie Foong, Bryan Veal, and Dan Williams.
\newblock {Platform Storage Performance with 3D XPoint Technology}.
\newblock {\em Proceedings of the IEEE}, 105(9):1822--1833, 2017.

\bibitem{helland1987group}
Pat Helland, Harald Sammer, Jim Lyon, Richard Carr, Phil Garrett, and Andreas
  Reuter.
\newblock {Group Commit Timers and High Volume Transaction Systems}.
\newblock In {\em International Workshop on High Performance Transaction
  Systems}, pages 301--329, 1987.

\bibitem{huang2018forca}
Haixin Huang, Kaixin Huang, Litong You, and Linpeng Huang.
\newblock {Forca: Fast and Atomic Remote Direct Access to Persistent Memory}.
\newblock In {\em ICCD}, pages 246--249, 2018.

\bibitem{huang2014nvram}
Jian Huang, Karsten Schwan, and Moinuddin~K Qureshi.
\newblock {NVRAM-aware Logging in Transaction Systems}.
\newblock {\em Proceedings of the VLDB Endowment}, 8(4):389--400, 2014.

\bibitem{ib}
{InfiniBand Trade Association}.
\newblock \url{http://www.infinibandta.org} (accessed Sep. 2021), 1999.

\bibitem{pmem-atomicity}
{Intel}.
\newblock {{PMEM Write Atomicity}}.
\newblock
  \url{https://software.intel.com/content/www/us/en/develop/articles/persistent-memory-replication-over-traditional-rdma-part-1-understanding-remote-persistent.html#inpage-nav-4-2}
  (accessed Sep. 2021).

\bibitem{vtune}
{Intel}.
\newblock {Intel VTune Profiler}.
\newblock
  \url{https://software.intel.com/content/www/us/en/develop/tools/vtune-profiler.html}
  (accessed Sep. 2021), 2014.

\bibitem{pmdk}
Intel.
\newblock {PMDK}.
\newblock \url{https://github.com/pmem/pmdk} (accessed Sep. 2021), 2014.

\bibitem{pcm}
{Intel}.
\newblock {Processor Counter Monitor}.
\newblock \url{https://github.com/opcm/pcm} (accessed Sep. 2021), 2017.

\bibitem{pmem-blog}
{Intel}.
\newblock {300 nanoseconds}.
\newblock \url{https://pmem.io/2019/12/19/performance.html} (accessed Sep.
  2021), 2019.

\bibitem{islam2016high}
Nusrat~Sharmin Islam, Md~Wasi-ur Rahman, Xiaoyi Lu, and Dhabaleswar~K Panda.
\newblock {High Performance Design for HDFS with Byte-Addressability of NVM and
  RDMA}.
\newblock In {\em ICS}, page~8, 2016.

\bibitem{kim2014evaluating}
Hyojun Kim, Sangeetha Seshadri, Clement~L Dickey, and Lawrence Chiu.
\newblock {Evaluating Phase Change Memory for Enterprise Storage Systems: A
  Study of Caching and Tiering Approaches}.
\newblock In {\em FAST}, pages 33--45, 2014.

\bibitem{kim2016nvwal}
Wook-Hee Kim, Jinwoong Kim, Woongki Baek, Beomseok Nam, and Youjip Won.
\newblock {NVWAL: Exploiting NVRAM in Write-Ahead Logging}.
\newblock In {\em ASPLOS}, pages 385--398, 2016.

\bibitem{lee2009architecting}
Benjamin~C Lee, Engin Ipek, Onur Mutlu, and Doug Burger.
\newblock {Architecting Phase Change Memory as a Scalable DRAM Alternative}.
\newblock In {\em ISCA}, pages 2--13, 2009.

\bibitem{lee2010phase}
Benjamin~C Lee, Ping Zhou, Jun Yang, Youtao Zhang, Bo~Zhao, Engin Ipek, Onur
  Mutlu, and Doug Burger.
\newblock {Phase-Change Technology and the Future of Main Memory}.
\newblock {\em IEEE Micro}, 30(1):143, 2010.

\bibitem{li2020rdmp}
Tianxi Li, Dipti Shankar, Shashank Gugnani, and Xiaoyi Lu.
\newblock {RDMP-KV: Designing Remote Direct Memory Persistence based Key-Value
  Stores with PMEM}.
\newblock In {\em SC}, pages 722--735, 2020.

\bibitem{lu2017octopus}
Youyou Lu, Jiwu Shu, Youmin Chen, and Tao Li.
\newblock {Octopus: An RDMA-enabled Distributed Persistent Memory File System}.
\newblock In {\em USENIX ATC}, pages 773--785, 2017.

\bibitem{ma2020asymnvm}
Teng Ma, Mingxing Zhang, Kang Chen, Zhuo Song, Yongwei Wu, and Xuehai Qian.
\newblock {AsymNVM: An Efficient Framework for Implementing Persistent Data
  Structures on Asymmetric NVM Architecture}.
\newblock In {\em ASPLOS}, pages 757--773, 2020.

\bibitem{mao2012cache}
Yandong Mao, Eddie Kohler, and Robert~Tappan Morris.
\newblock {Cache Craftiness for Fast Multicore Key-Value Storage}.
\newblock In {\em EuroSys}, pages 183--196, 2012.

\bibitem{mittal2016survey}
Sparsh Mittal and Jeffrey~S Vetter.
\newblock {A Survey of Software Techniques for Using Non-Volatile Memories for
  Storage and Main Memory Systems}.
\newblock {\em IEEE Transactions on Parallel and Distributed Systems},
  27(5):1537--1550, 2016.

\bibitem{Mohan1992ARIESAT}
C.~Mohan, Donald~J. Haderle, Bruce~G. Lindsay, Hamid Pirahesh, and Peter~M.
  Schwarz.
\newblock {ARIES: A Transaction Recovery Method Supporting Fine-Granularity
  Locking and Partial Rollbacks Using Write-Ahead Logging}.
\newblock {\em ACM Transactions on Database Systems (TODS)}, 17:94--162, 1992.

\bibitem{nalli2017analysis}
Sanketh Nalli, Swapnil Haria, Mark~D Hill, Michael~M Swift, Haris Volos, and
  Kimberly Keeton.
\newblock {An Analysis of Persistent Memory use with WHISPER}.
\newblock In {\em ASPLOS}, pages 135--148, 2017.

\bibitem{pmem-rocksdb}
{Peifeng Si}.
\newblock {Persistent Memory Storage Engine for RocksDB}.
\newblock {https://github.com/pmem/pmem-rocksdb} (accessed Sep. 2021), 2019.

\bibitem{shan2017distributed}
Yizhou Shan, Shin-Yeh Tsai, and Yiying Zhang.
\newblock {Distributed Shared Persistent Memory}.
\newblock In {\em SoCC}, pages 323--337, 2017.

\bibitem{pmem-rel}
{Steve Swanson}.
\newblock {Engineering Reliable Persistence}.
\newblock \url{https://www.sigarch.org/engineering-reliable-persistence/}
  (accessed Sep. 2021), 2017.

\bibitem{memristor}
Dmitri~B Strukov, Gregory~S Snider, Duncan~R Stewart, and R~Stanley Williams.
\newblock {The Missing Memristor Found}.
\newblock {\em Nature}, 453(7191):80, 2008.

\bibitem{taleb2018tailwind}
Yacine Taleb, Ryan Stutsman, Gabriel Antoniu, and Toni Cortes.
\newblock {Tailwind: Fast and Atomic RDMA-based Replication}.
\newblock In {\em USENIX ATC}, pages 851--863, 2018.

\bibitem{rcommit}
Tom Talpey.
\newblock {Remote Access to Ultra-Low-Latency Storage - SNIA}.
\newblock
  \url{https://www.snia.org/sites/default/files/SDC15_presentations/persistant_mem/Talpey-Remote_Access_Storage.pdf},
  2015.

\bibitem{tsai2020disaggregating}
Shin-Yeh Tsai, Yizhou Shan, and Yiying Zhang.
\newblock {Disaggregating Persistent Memory and Controlling Them Remotely: An
  Exploration of Passive Disaggregated Key-Value Stores}.
\newblock In {\em USENIX ATC}, pages 33--48, 2020.

\bibitem{tulapurkar2005spin}
AA~Tulapurkar, Y~Suzuki, A~Fukushima, H~Kubota, H~Maehara, K~Tsunekawa,
  DD~Djayaprawira, N~Watanabe, and S~Yuasa.
\newblock {Spin-Torque Diode Effect in Magnetic Tunnel Junctions}.
\newblock {\em Nature}, 438(7066):339, 2005.

\bibitem{tweedie1998journaling}
Stephen~C Tweedie.
\newblock {Journaling the Linux ext2fs Filesystem}.
\newblock In {\em The Fourth Annual Linux Expo}, 1998.

\bibitem{wang2015building}
Guozhang Wang, Joel Koshy, Sriram Subramanian, Kartik Paramasivam, Mammad
  Zadeh, Neha Narkhede, Jun Rao, Jay Kreps, and Joe Stein.
\newblock {Building a Replicated Logging System with Apache Kafka}.
\newblock {\em Proceedings of the VLDB Endowment}, 8(12):1654--1655, 2015.

\bibitem{wang2017query}
Tianzheng Wang, Ryan Johnson, and Ippokratis Pandis.
\newblock {Query Fresh: Log Shipping on Steroids}.
\newblock {\em Proceedings of the VLDB Endowment}, 11(4):406--419, 2017.

\bibitem{xu2019finding}
Jian Xu, Juno Kim, Amirsaman Memaripour, and Steven Swanson.
\newblock {Finding and Fixing Performance Pathologies in Persistent Memory
  Software Stacks}.
\newblock In {\em ASPLOS}, pages 427--439, 2019.

\bibitem{yang2019orion}
Jian Yang, Joseph Izraelevitz, and Steven Swanson.
\newblock {Orion: A Distributed File System for Non-Volatile Main Memory and
  RDMA-Capable Networks}.
\newblock In {\em FAST}, pages 221--234, 2019.

\bibitem{yang2020filemr}
Jian Yang, Joseph Izraelevitz, and Steven Swanson.
\newblock {FileMR: Rethinking RDMA Networking for Scalable Persistent Memory}.
\newblock In {\em NSDI}, pages 111--125, 2020.

\bibitem{yang-fast20}
Jian Yang, Juno Kim, Morteza Hoseinzadeh, Joseph Izraelevitz, and Steve
  Swanson.
\newblock {An Empirical Guide to the Behavior and Use of Scalable Persistent
  Memory}.
\newblock In {\em FAST}, pages 169--182, 2020.

\bibitem{zhang2019pangolin}
Lu~Zhang and Steven Swanson.
\newblock {Pangolin: A Fault-Tolerant Persistent Memory Programming Library}.
\newblock In {\em USENIX ATC}, pages 897--912, 2019.

\bibitem{zhang2015mojim}
Yiying Zhang, Jian Yang, Amirsaman Memaripour, and Steven Swanson.
\newblock {Mojim: A Reliable and Highly-Available Non-Volatile Memory System}.
\newblock In {\em ASPLOS}, pages 3--18, 2015.

\end{thebibliography}
